\documentclass[nofootinbib, preprintnumbers, superscriptaddress,notitlepage]{revtex4-1}

\usepackage{amsmath,amssymb,amscd,simplewick,braket,verbatim}
\usepackage{listings}
\usepackage{dsfont}
\usepackage{slashed}
\usepackage{color}
\usepackage{ulem}
\usepackage{float}
\usepackage{slashbox}

\allowdisplaybreaks[4]

\usepackage{graphicx}
\usepackage{epstopdf}
\usepackage{subfigure}
\usepackage{epsfig}
\usepackage{xcolor}
\usepackage[colorlinks=true,
            linkcolor=blue,
            urlcolor=blue,
            citecolor=green,          
            bookmarks=true,
            bookmarksnumbered=true,
            breaklinks=true,
            pdfpagemode=Fullscreen,
            pdfstartview=FitBH]{hyperref}

\hypersetup{pdfauthor = {Shao-Feng Ge},
	     pdftitle = {}, 
	     pdfsubject = {}, 
             pdfkeywords = {}, 
	     pdfcreator = {LaTeX with hyperref package},
	     pdfproducer = {dvips + ps2pdf} }

\definecolor{gesfpurple}{rgb}{0.47,0.19,0.42}

\definecolor{gesflanse}{rgb}{0.00,0.50,0.50}

\definecolor{gesfblue}{rgb}{0.08,0.42,0.76}

\definecolor{gesfred}{rgb}{1,0,0}

\definecolor{gesfwhite}{rgb}{1,1,1}

\definecolor{gesfblack}{rgb}{0,0,0}

\newcommand{\tr}{\mbox{Tr}}
\definecolor{OliveGreen}{cmyk}{0.64,0,0.95,0.40}

\newcommand{\gsec}[1]{{\hypersetup{linkcolor=red}Sec.~\ref{#1}\hypersetup{linkcolor=blue}}}
\newcommand{\gapp}[1]{{\hypersetup{linkcolor=red}App.~\ref{#1}\hypersetup{linkcolor=blue}}}
\newcommand{\geqn}[1]{\hypersetup{linkcolor=blue}(\ref{#1})\hypersetup{linkcolor=blue}}
\newcommand{\gfig}[1]{{\hypersetup{linkcolor=violet}Fig.~\ref{#1}\hypersetup{linkcolor=blue}}}
\newcommand{\gtab}[1]{{\hypersetup{linkcolor=gesflanse}Tab.~\ref{#1}\hypersetup{linkcolor=blue}}}

\graphicspath{{figs/}}

\usepackage{tikz} 
\usepackage{tkz-euclide}
\usetikzlibrary{backgrounds}
\usetikzlibrary{decorations.pathmorphing}
\usetikzlibrary{arrows.meta}
\tikzset{
    v/.style={decorate, decoration={snake, segment length=2.mm, amplitude=0.5mm}, draw},
    f/.style={draw,decoration={markings,mark=at position #1 with {\arrow[]{Latex[length=1.5mm,width=1.5mm]}}},postaction={decorate},node contents=#1},
    fr/.style 2 args={draw,decoration={markings,mark=at position #1 with {\arrow[rotate=#2]{Latex[length=1.5mm,width=1.5mm]}}},postaction={decorate},node contents=#1},
    f/.default=.6,
    fr/.default={.6}{0},
    fb/.style={draw,decoration={markings,mark=at position #1 with {\arrowreversed[]{Latex[length=1.5mm,width=1.5mm]}}},postaction={decorate},node contents=#1},
    fb/.default=.4,
    fnar/.style={draw},
    g/.style={decorate, draw,  decoration={coil,amplitude=3pt, segment length=3.5pt}},
    s/.style={dashed,draw, postaction={decorate},
        decoration={markings,mark=at position .55 with {\arrow[very thick]{latex}}}},
    sb/.style={dashed,draw, postaction={decorate},
        decoration={markings,mark=at position .55 with {\arrowreversed[draw=black,very thick]{latex}}}},
    snar/.style={dashed,draw,line width =1.25pt},
}

\tikzset{mystyle/.style={line width=1,baseline,scale=0.6, every node/.style={scale=0.6}}}
\tikzset{circlestyle/.style={preaction={fill=white},postaction={pattern=north west lines},fill=cyan,fill opacity=1,draw=black}}
\tikzset{circlestyle2/.style={preaction={fill=white},postaction={pattern=north east lines},fill=red,fill opacity=0.5,draw=black}}

\begin{document}
\fontsize{12pt}{14pt}\selectfont

\title{\Large Revisiting the Fermionic Dark Matter Absorption on Electron Target}

\author{Shao-Feng Ge}
\email{gesf@sjtu.edu.cn}
\affiliation{Tsung-Dao Lee Institute (TDLI) \& School of Physics and Astronomy (SPA),
Shanghai Jiao Tong University (SJTU), Shanghai 200240, China}
\affiliation{Key Laboratory for Particle Astrophysics and Cosmology (MOE) \& Shanghai Key Laboratory for Particle Physics and Cosmology, Shanghai Jiao Tong University, Shanghai 200240, China}

\author{Xiao-Gang He}
\email{hexg@phys.ntu.edu.tw}
\affiliation{Tsung-Dao Lee Institute (TDLI) \& School of Physics and Astronomy (SPA),
Shanghai Jiao Tong University (SJTU), Shanghai 200240, China}
\affiliation{Department of Physics, National Taiwan University, Taipei 10617}

\author{Xiao-Dong Ma}
\email{maxid@sjtu.edu.cn}
\affiliation{Tsung-Dao Lee Institute (TDLI) \& School of Physics and Astronomy (SPA),
Shanghai Jiao Tong University (SJTU), Shanghai 200240, China}
\affiliation{Key Laboratory for Particle Astrophysics and Cosmology (MOE) \& Shanghai Key Laboratory for Particle Physics and Cosmology, Shanghai Jiao Tong University, Shanghai 200240, China}

\author{Jie Sheng}
\email{shengjie04@sjtu.edu.cn}
\affiliation{Tsung-Dao Lee Institute (TDLI) \& School of Physics and Astronomy (SPA),
Shanghai Jiao Tong University (SJTU), Shanghai 200240, China}
\affiliation{Key Laboratory for Particle Astrophysics and Cosmology (MOE) \& Shanghai Key Laboratory for Particle Physics and Cosmology, Shanghai Jiao Tong University, Shanghai 200240, China}

\begin{abstract}
\fontsize{10pt}{12pt}\selectfont
We perform a systematic study of the fermionic DM absorption
interactions on electron target in the context of effective
field theory. The fermionic DM absorption is not just sensitive
to sub-MeV DM with efficient energy release, but also gives a
unique signature with clear peak in the electron recoil
spectrum whose shape is largely determined by the atomic effects.
Fitting with the Xenon1T and PandaX-II data prefers DM mass at
$m_\chi = 59$\,keV and 105\,keV, respectively, while the cut-off
scale is probed up to around 1\,TeV.
The DM overproduction in the early Universe,
the invisible decay effect on the cosmological evolution,
and the astrophysical X(gamma)-ray from the DM visible decays
are thoroughly explored to give up-to-date constraints. With
stringent bounds on the tensor and pseudo-scalar operators,
the other fermionic DM operators are of particular interest at
tonne-scale direct
detection experiments such as PandaX-4T, XENONnT, and LZ.
\end{abstract}

\maketitle

\section{Introduction}

The nature of dark matter (DM) remains a mysterious puzzle in
our understanding of the Universe \cite{Young:2016ala, Arbey:2021gdg}.
The possible particle characteristic of DM is a well-motivated scenario
to be probed by the direct detection experiments
\cite{Liu:2017drf,Billard:2021uyg} and indirect observations 
\cite{Leane:2020liq, Slatyer:2021qgc}.
The stability of DM particle is
usually realized by some discrete symmetry such as
$\mathbb Z_2$ \cite{Albert:2016osu, Arun:2017uaw, Lin:2019uvt}.
A direct consequence is
that in direct detection, the scattering process has a
DM particle in the initial state and another DM particle
in the final one. The energy deposit comes from the DM
kinetic energy. With typical experimental threshold at
$\mathcal O(1)$\,keV, direct detection experiments are
only sensitive to the DM mass above GeV. In the GeV$\sim$TeV
mass range, the null result from the direct detection experiments
has put very strong limit on the DM interaction strength with
the Standard Model (SM) particles
\cite{Roszkowski:2017nbc, Bottaro:2021snn}. In contrast,
the cross section of sub-GeV light DM scattering
with SM particles is much less stringently constrained
and can still be large \cite{Liu:2017drf,Billard:2021uyg}.
More attention has been turned to light DM alternatives
with sub-GeV mass \cite{Davis:2015vla} or even lighter ones
such as the sterile neutrino DM \cite{Drewes:2016upu,Abazajian:2017tcc,
Boyarsky:2018tvu,Kopp:2021jlk}.

However, one difficulty
for the light DM detection is its small recoil energy.
For a typical DM scattering with nuclei target, the recoil
energy $T_r = 4 m_\chi m_A T_\chi / (m_\chi + m_A)^2$ is
proportionally scaled from the DM kinetic energy $T_\chi$.
The most efficient energy transfer happens when the DM
mass $m_\chi$ is roughly the size of the atomic mass $m_A$
of the nuclei target, $T_r \approx T_\chi$.
With light DM, $m_\chi \ll m_A$, the recoil energy
$T_r \approx 4 m_\chi T_\chi / m_A$ decreases with not just
the DM mass $m_\chi$ but also its kinetic energy $T_\chi$.
More importantly, the DM kinetic energy is also proportional
to its mass, $T_\chi \approx \frac 1 2 m_\chi v^2_\chi$,
while the distribution of its velocity $v_\chi$ is fixed by
the galaxy gravitational potential \cite{Bozorgnia:2016ogo}.
Altogether, the nuclear
recoil from the light DM scattering scales with $m^2_\chi$.
This explains why the direction detection sensitivity
deteriorate fast in the sub-GeV range. It is desirable to
find possible ways to overcome this difficulty.

There are several ways of improving the detection of light
DM. For nuclear recoil, the detection threshold can be lowered
by using Germanium point-contact detector \cite{Collar:2021fcl},
bolometer \cite{Abdelhameed:2019hmk,Pirro:2017ecr},
nuclear bremsstrahlung
\cite{Kouvaris:2016afs,GrillidiCortona:2020owp}, and Migdal effect
\cite{Ibe:2017yqa, Baxter:2019pnz, Essig:2019xkx,Flambaum:2020xxo,GrillidiCortona:2020owp,Wang:2021oha,Acevedo:2021kly,Bell:2021zkr,Knapen:2020aky,Nakamura:2020kex,Liu:2020pat,Dey:2020sai}.
Or one may replace the nuclear recoil by electron recoil.
Then, the elastic recoil energy is
$T_r = 4 m_\chi m_e T_\chi / (m_\chi + m_e)^2$
which removes the suppression factor $m_\chi/m_A$.
In addition to using the conventional detector for measuring
the electron recoils, various new technology has been developed.
From the condensed matter side, the typically small energy gap
is of great advantage to build a low threshold detector such as
using superheated liquid \cite{SH},
super-conduction \cite{SC}, Fermi degenerate materials \cite{FD},
super-fluid \cite{SF}, scintillation \cite{Scintillator},
magnetic molecular \cite{Molecular}, Dirac material
\cite{DiracMaterial}, diamond crystal \cite{Diamond},
nanowire \cite{nanoWire}, nanotube \cite{Cavoto:2017otc},
magnon \cite{magnon}, graphene \cite{Graphene},
and plasmon \cite{Plasmon}. In particular, a semi-conductor
detector such as skipper CCD is very sensitive to single
electron events \cite{SkipperCCD}. The biological DNA
also provides an interesting possibility \cite{DNA}.

On the other hand, DM particles upscattered to higher
energy can also overcome the detection threshold. Several
possibilities have been discussed in the literature.
The nonrelativistic DM particles can be boosted
by the cosmic rays to gain sufficient energy
\cite{Cappiello:2018hsu,Bringmann:2018cvk,Ema:2018bih,
Dent:2019krz,Wang:2019jtk,Ge:2020yuf, Lei:2020mii,
Xia:2020apm,Feng:2021hyz,Chen:2021ifo,Xia:2021vbz,
Dent:2020syp,Bell:2021xff,Wang:2021nbf,Cho:2020mnc,
Cao:2020bwd}.
This cosmic ray boosted DM (CRDM) scenario can happen as long
as DM interacts with SM particles which is exactly the
foundation of DM direct detection. Actual experimental
search with real DM direct detection data has been carried
out by PandaX-II \cite{PandaX-II:2021kai} and CDEX
\cite{CDEX:2022fig}, in addition to those constraints
from neutrino experiments
\cite{Ema:2018bih,Cappiello:2019qsw,Guo:2020drq,Ema:2020ulo,
PROSPECT:2021awi,Chauhan:2021fzu} and indirect
constraints \cite{Cappiello:2018hsu,Guo:2020oum}.
The CRDM may also be
produced by astrophysical neutrinos
\cite{Pandey:2018wvh,Zhang:2020nis,Jho:2021rmn,Das:2021lcr,
Chao:2021orr,Ghosh:2021vkt} and blazar \cite{Wang:2021jic}.
If the DM particle is light
enough, it is also possible for them to be produced by
the cosmic ray interactions with the atmosphere
\cite{Alvey:2019zaa, Plestid:2020kdm,Kachelriess:2021man}.
Another place to boost light DM is the Sun
\cite{Kouvaris:2015nsa,An:2017ojc,Emken:2017hnp,Emken:2021lgc,
Chen:2020gcl,An:2021qdl}. With multiple components, the
boosted light DM can also happen inside the dark sector
\cite{Agashe:2014yua,Berger:2014sqa,Cherry:2015oca,Fornal:2020npv,
Chen:2020oft}.

Another possibility is the fermionic DM absorption.
The upscattered DM scenarios mentioned in the above paragraph
can probe light DM, but the dependence on the DM mass may not
be significant. This is because smaller mass usually means smaller
effect on the kinematics, especially if DM particles are highly
boosted. In order for DM detection to be
sensitive to the light DM mass, the mass term should dominate
the relevant kinematics. Namely, non-relativistic DM may have
some advantage in this regard. If a nonrelativistic DM releases
all its mass into energy, its mass is the dominant factor and
the detection threshold can also be overcome with efficient
amplification by the speed of light, $E = m c^2$. This is exactly
the idea of DM absorption for bosonic
\cite{Pospelov:2008jk, An:2014twa, Bloch:2016sjj,Hochberg:2016ajh,
Hochberg:2016sqx,Green:2017ybv,Arvanitaki:2017nhi,vonKrosigk:2020udi,
Mitridate:2021ctr} and fermionic
\cite{Dror:2019onn,Dror:2019dib,Dror:2020czw} DM.

This paper is organized as follows. 
In \gsec{sec:FADM}, we introduce the motivation for the sub-MeV 
fermionic absorption DM and enumerate all the possible effective
absorption operators. In \gsec{sec:dd} we discuss the signal in
direct detection experiment. Then we evaluate the constraints
from DM overproduction of the early Universe in
\gsec{sec:overproduction} as well as the cosmological evolution
constraint on the invisible decay $\chi \rightarrow 3 \nu$
and astrophysical constraints with X(gamma)-rays on the visible
decay modes $\chi \rightarrow \nu + \gamma (s)$ in
\gsec{sec:constraints}. More details about the calculation of
DM decay is provided in \gsec{sec:DMDC}. Our main results are
summarized in \gsec{sec:summary}. On the technical side, we
provide simplified algorithms for a general-purpose analytic
$\chi^2$ fit with collective marginalization in \gapp{app:chi2fit}.

\section{Sub-MeV Fermionic Absorption DM on Electron Target}
\label{sec:FADM}

As pointed out above, the light DM has intrinsic difficulty
in the direct detection experiments due to energy threshold.
One possible way of overcoming this comes from the fermionic
DM absorption, $\chi e \rightarrow \nu e$, where the DM particle
$\chi$ scatters into a massless SM neutrino $\nu$. Placing
neutrino in the final state not only conserves charge but
also is the most economical choice to
maximize the energy release. Then the DM mass $m_\chi$ is
wholy converted to the electron recoil and neutrino energies.
The fermionic DM absorption on a nuclei target is also possible
\cite{Dror:2019onn,Dror:2019dib} but requires heavier DM above MeV mass to
overcome the detection threshold. In our current paper, we
focus on the electron target that is optimal for sub-MeV DM
\cite{Dror:2020czw}.

For a free electron target at rest, the electron recoil energy
is $T_r \approx m^2_\chi / 2 m_e$ \cite{Dror:2020czw},
which is a good approximation for $m_\chi \ll m_e$.
A keV scale DM can already produce large enough electron recoil
energy to overcome the detection threshold that is typically
1\,keV for the electron signal \cite{XENON:2020rca,PandaX-II:2021nsg}.
Although larger electron recoil energy $T_r$ is better
for direct detection threshold, the DM mass is not larger
the better. For $m_\chi = 1$\,MeV, the electron recoil
approaches 1\,MeV which may saturate the detector capability.
So we focus on the sub-MeV DM mass range with
1\,keV$\lesssim m_\chi \lesssim 1$\,MeV across this paper.

To make a systematic study of the fermionic DM absorption,
we take the effective field theory (EFT)
approach for a model independent analysis. As argued at
the beginning of this section, the relevant degrees of
freedom are the light SM particles, electron and neutrino,
augmented with an additional DM particle. Usually, the
SM gauge symmetries $U(1)_Y \times SU(2)_L \times SU(3)_c$
is kept intact for an EFT approach. Nevertheless,
the DM direct detection happens at low energy where the
electroweak part is broken. Only the electromagnetic
$U(1)_{\rm em}$ is a good symmetry to guide the construction
of EFT operators as far as gauge symmetry of the theory
is concerned. The strong interaction $SU(3)_c$ is of
no relevance since no color degrees of freedom are involved.

For the fermionic DM absorption on the electron target,
the leading local interactions are dimension-six operators
involving a dark matter particle $\chi$, an active SM
neutrino $\nu$ and an electron current,
\begin{subequations}
\begin{eqnarray}
&&{\cal O}_{e\nu\chi}^{S} \equiv (\bar ee)(\bar\nu_L \chi_R),
\\
&&{\cal O}_{e\nu\chi}^{P} \equiv (\bar e i\gamma_{5}e)(\bar\nu_L\chi_R),
\\
&&{\cal O}_{e\nu\chi}^{V} \equiv (\bar e\gamma_{\mu}e)(\bar\nu_L\gamma^{\mu} \chi_L),
\\
&&{\cal O}_{e\nu\chi}^{A} \equiv (\bar e\gamma_{\mu}\gamma_{5}e)(\bar\nu_L\gamma^{\mu} \chi_L),
\\
&&{\cal O}_{e\nu\chi}^{T} \equiv (\bar e\sigma_{\mu\nu}e)(\bar\nu_L\sigma^{\mu\nu}\chi_R),
\end{eqnarray}
\label{eq:operator}
\end{subequations}
and their Hermitian conjugates. For completeness, we have
considered all the five independent Lorentz structures for
the electron bilinear (scalar [S], pseudo-scalar [P], vector [V],
axial-vector [A] and tensor [T]). The neutrino field is
taken to be the SM left-handed component $\nu_L$ and the DM
$\chi$ is assumed to be a Dirac particle for convenience.
Any other types of operators can be converted to those in
\geqn{eq:operator} by Dirac gamma matrix identities and
Fierz transformations \cite{Nieves:2003in,Nishi04,Liao:2012uj}.
For instance, the operator
$(\bar e\sigma_{\mu\nu}\gamma_5 e)(\bar\nu_L\sigma^{\mu\nu}\chi_R)$ 
is equivalent to ${\cal O}_{e\nu\chi}^{T}$ by the identities 
$\sigma^{\mu\nu}\gamma^5={i \over 2}
\epsilon^{\mu\nu\rho\sigma}\sigma_{\rho\sigma}$ and $\gamma_5 
P_R=P_R$ with $P_R$ being the right-handed projection operator. 

Each operator carries a Wilson coefficient $C_i$. Since the
operators are already dimension 6, $C_i \equiv 1 / \Lambda^2$
carry two units of inverse mass dimension, [mass]$^{-2}$.
Given new physics scenario, a heavy mediator can be integrated
away to match the effective interaction. Equivalently,
the effective scale $\Lambda$ can be identified with the
heavy mediator mass up to some dimensionless coupling constants.  
Although there is no fundamental principle to forbidden the
above operators to simultaneously appear, we consider them
separately in the following discussions.

\subsection{Fermionic DM Absorption on Electron and Atomic Effects}
\label{sec:dd}

The local DM distribution around the Sun is roughly known
from the hydrodynamic simulation of our Milky Way galaxy
\cite{Bozorgnia:2016ogo}. With the DM energy density,
$\rho_\chi \approx 0.4$\,GeV/cm$^3$, its number density
$n_\chi = \rho_\chi / m_\chi$ is inversely proportional to
mass $m_\chi$.
In addition, the DM velocity distribution is determined by
the gravitational potential of the galaxy matter and dark
matter. Around the Sun, the DM velocity follows the
Maxwell-Boltzmann distribution and peaks around typically
$v_\chi \sim (200 \sim 300)$\,km/s.
In other words, the DM in our solar system is non-relativistic
with only $\mathcal O(10^{-3})$ of the speed
of light. Since the absorption process releases the whole
DM mass as the energy of final-state particles, its kinetic
energy $T_\chi \approx \frac 1 2 m_\chi v^2_\chi$ is
negligibly small, $T_\chi / m_\chi \sim \mathcal O(10^{-6})$.
We can take both the initial electron and DM particle as at
rest to estimate the scattering cross section $\sigma_{\chi e}$,
\begin{subequations}
\begin{eqnarray}
  \sigma^S_{\chi e} v_\chi
& \approx & 
  \frac 1 {\Lambda^4}
  \frac {m_\chi^2 (2m_e+m_\chi)^4}{64\pi (m_e+m_\chi)^4},
\\
  \sigma^P_{\chi e} v_\chi
& \approx & 
  \frac 1 {\Lambda^4}
  \frac {m_\chi^4 (2m_e+m_\chi)^2}{64\pi (m_e+m_\chi)^4},
\\
  \sigma^V_{\chi e} v_\chi
& \approx & 
  \frac 1 {\Lambda^4}
  { m_\chi^2 (2m_e+m_\chi)^2 (2m_e^2 + 4 m_e m_\chi + 3m_\chi^2 ) \over 32\pi (m_e+m_\chi)^4},
\\
  \sigma^A_{\chi e} v_\chi
& \approx & 
  \frac 1 {\Lambda^4}
  { m_\chi^2 (2m_e+m_\chi)^2 (6 m_e^2 + 8 m_e m_\chi + 3m_\chi^2 ) \over 32\pi (m_e+m_\chi)^4},
\\
  \sigma^T_{\chi e} v_\chi
& \approx & 
  \frac 1 {\Lambda^4}
  { m_\chi^2 (2m_e+m_\chi)^2 (6 m_e^2 + 10 m_e m_\chi + 5 m_\chi^2 ) \over 8\pi (m_e+m_\chi)^4 }. 
\end{eqnarray}
\label{eq:sigmachie}
\end{subequations}
Those terms suppressed by the DM velocity $v_\chi$ have
been neglected for clarity. Nevertheless, the DM velocity does
not disappear completely since the quantity that enters the
DM event rate is $\sigma_{\chi e} v_\chi$ as a whole.
In the limit of tiny DM mass, $m_\chi \ll m_e$, all cross
sections reduce to a universal value, $m^2_\chi / 4 \pi \Lambda^4$
\cite{Dror:2020czw}.

In reality, the initial and final electrons are subject
to the Coulomb potential produced by the central nuclei
and other surrounding electrons. There is no way for the
initial electron that is confined inside the atom to be
at rest. Not to say for a typical $\mathcal O($keV)
electron recoil, its kinetic energy is of the same size
as the binding energy. The atomic binding effect could
be large enough to affect the direct detection event rate.
We follow the general formalism developed with second
quantization for both the initial bound and final ionized
electron states \cite{Ge:2021snv}. The differential
cross section is then a convolution of the particle
scattering amplitude $|\mathcal M|^2$ and the atomic
$K$-factor $K_{nl} (T_r, |{\bf q}|)$,
\begin{eqnarray}
  \frac {d \sigma_{\chi e}}{d T_r} v_\chi
=
  \frac 1 {T_r} \sum_{nl} (4l +2)
  \int \frac{d^3 {\bf q}}{(2\pi)^3 2 E_\nu}
  \frac{1}{8 m_e^2 E_\chi}
  |\mathcal M|^2({\bf q}) K_{nl}(T_r, |{\bf q}|)
  (2\pi) \delta_E,
\label{eq:dSigmadTr1}
\end{eqnarray}
The summation over the principle quantum number $n$ and
the angular momentum $l$ is a product of the electron 
number $(2l + 1)$ of the state $\ket{nl}$ and the 
spin degree of freedom $2$.

The remaining
$\delta_E$ function comes from energy conservation. For
non-relativistic DM, its kinetic energy can be omitted
in comparsion with the energy release from the DM absorption
process,
\begin{eqnarray}
  \delta_E
\equiv
  \delta(m_\chi - |{\bf q}| - \Delta E_{nl}).
\label{eq:deltaE}
\end{eqnarray}
With vanishing mass,
the neutrino energy $E_\nu = |{\bf q}|$ is the same
as its momentum and equivalently the size of momentum transfer
$|{\bf q}|$. On the electron side, the energy gain,
$\Delta E_{nl} \equiv T_r - E_{nl}$, is the difference between the
ionized electron energy $T_r$ and the negative initial binding
energy $E_{nl}$ for the state $|nl\rangle$.
It is then desirable to first integrate
$|{\bf q}|$ away from the phase space element
$d^3 {\bf q} = |{\bf q}|^2 d |{\bf q}| d \Omega_q$. 
For the scattering with a bound electron, $|{\bf q}|$
and $T_r$ are usually independent variables due to the
unknown initial electron momentum inside an atom.
However, the energy conservation \geqn{eq:deltaE}
establishes a correlation, $|{\bf q}| = m_\chi - \Delta E_{nl}$.

On the other hand, the solid angle integration
$d \Omega_q \equiv d \cos \theta_q d \phi_q$ contains
the information of momentum transfer direction. However,
neither the $K$-factor $K_{nl}(T_r, |{\bf q}|)$ \cite{Ge:2021snv}
nor the scattering matrix element $|\mathcal M ({\bf q})|^2$
has dependence on the angular coordinates of ${\bf q}$.
For the latter, the angular independence happens due to
the fact that the DM velocity is negligible in the
absorption process. Without a preference, the direction
of ${\bf q}$ is not important either. The solid angle
integration then simply gives an overall $4 \pi$.
Consequently, \geqn{eq:dSigmadTr1} becomes,
\begin{eqnarray}
  \frac {d \langle \sigma_{\chi e} v_\chi \rangle}{d T_r}
=
  \sum_{nl} (4 l + 2) \frac 1 {T_r}
  \frac {m_\chi -  \Delta E_{nl}} {16 \pi m^2_e m_\chi}
  |\mathcal M|^2({\bf q}) K_{nl}(T_r, |{\bf q}|),
\label{eq:dSigmadTr2}
\end{eqnarray}
where we have implemented the facts that $E_\nu = |{\bf q}|$
and $|{\bf q}| = m_\chi - \Delta E_{nl}$. Since the left-hand
sides of \geqn{eq:dSigmadTr1} and \geqn{eq:dSigmadTr2}
are independent of the DM velocity, $\sigma_{\chi e} v_\chi$ is
essentially $\langle \sigma_{\chi e} v_\chi \rangle$.
Although there are five operators in \geqn{eq:operator},
the scattering matrix element $|\mathcal M|^2$ has only
two different forms,
\begin{eqnarray}
  \left| \mathcal M^{(S,V,A,T)} \right|^2
=
  (4, 4, 12, 48)
  \times \frac 1 {\Lambda^4} m_\chi m^2_e (m_\chi - \Delta E_{nl}),
\qquad
  \left| \mathcal M^P \right|^2
=
  \frac 1 {\Lambda^4} m_\chi (m_\chi - \Delta E_{nl})^3.
\label{eq:M2}
\end{eqnarray}
The pseudo-scalar (P) case is quite special since the
matrix element intrinsically has momentum transfer
dependence as elaborated in \cite{Ge:2021snv}. For
comparison, the others have exactly the same structure.

\begin{figure}[t]
\centering
\includegraphics[width=0.48\textwidth]{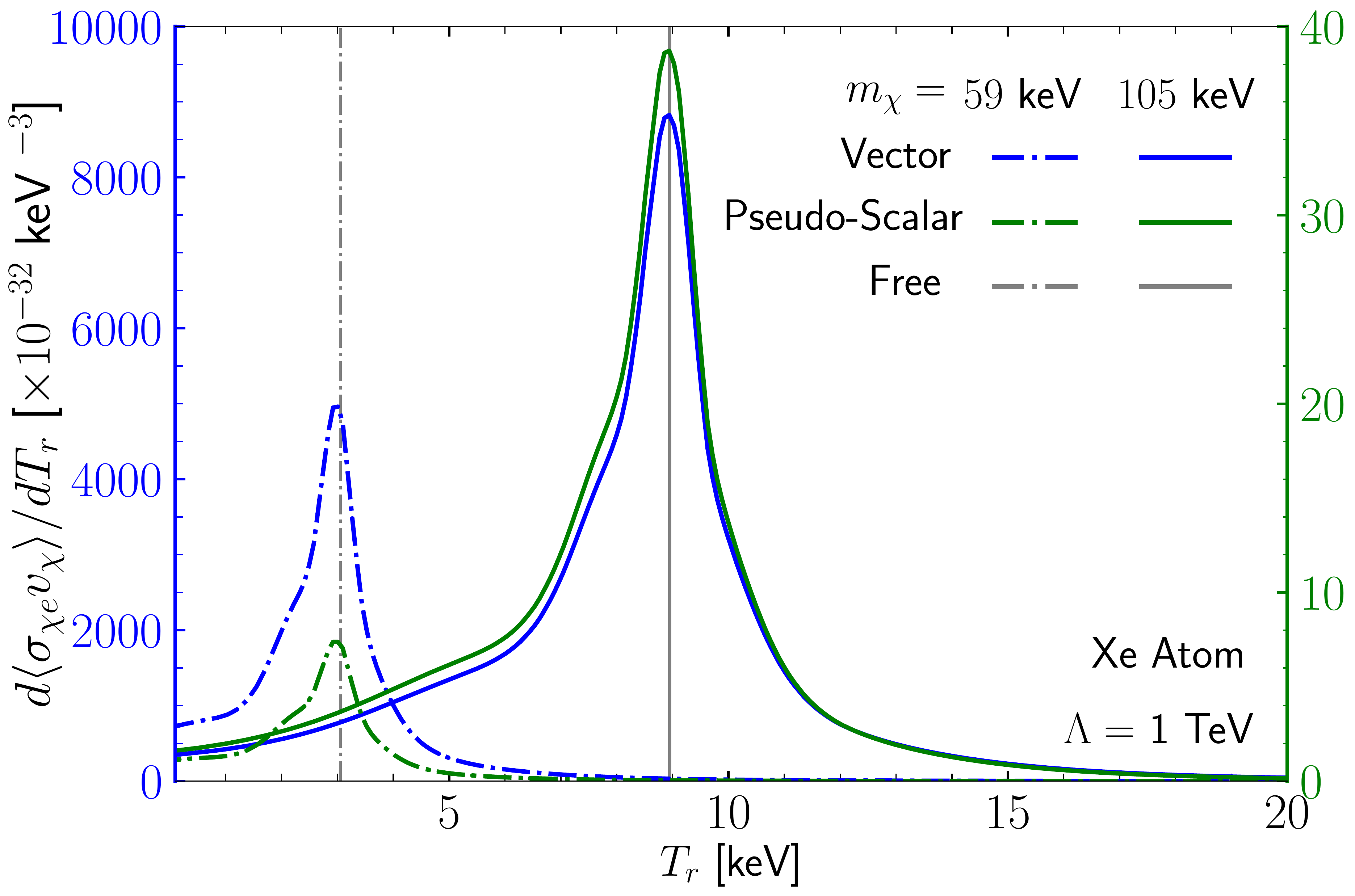}
\hfill
\includegraphics[width=0.48\textwidth]{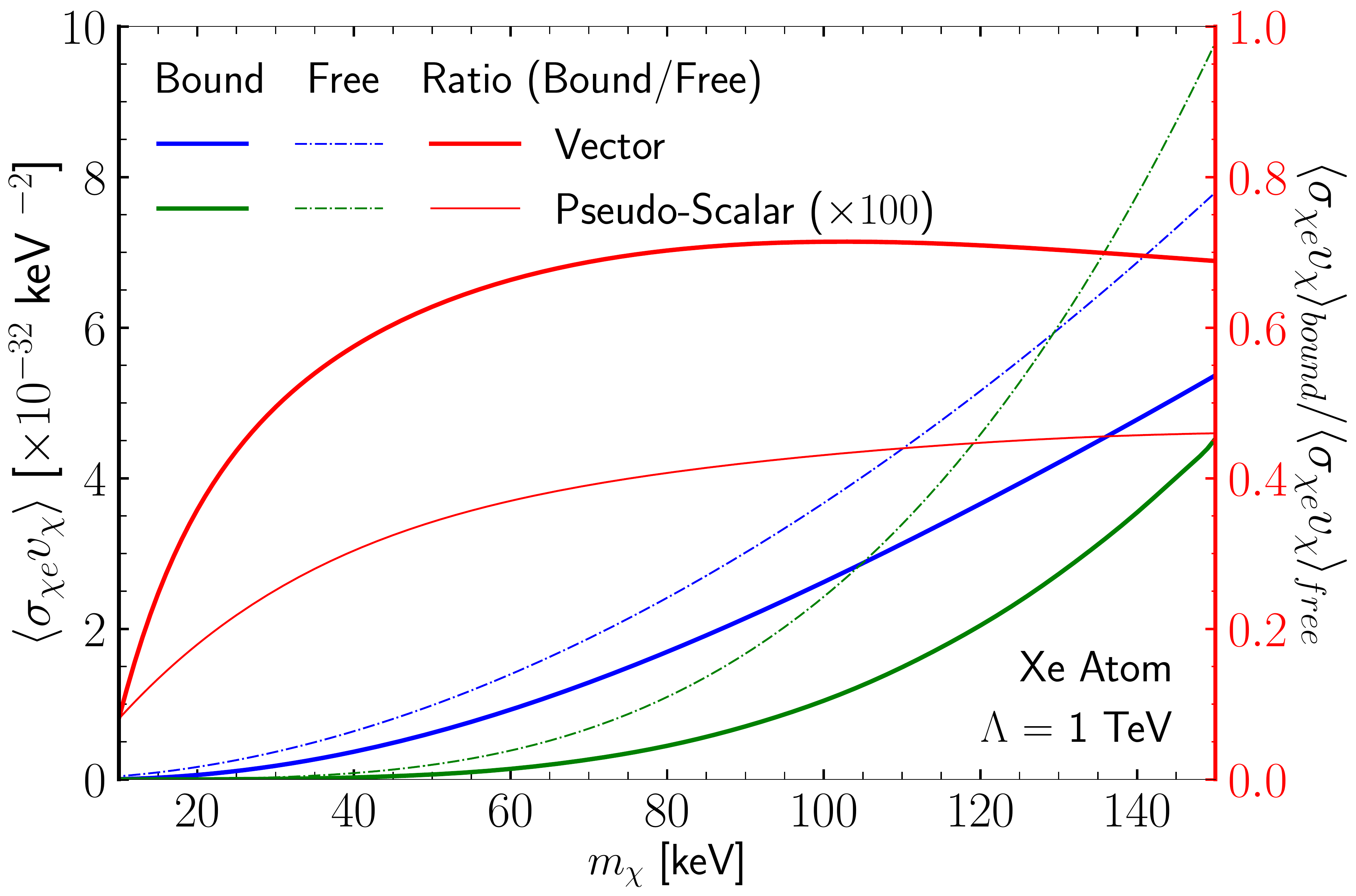}
\caption{({\bf Left}) The atomic effects of Xe on the electron
recoil spectrum with two typical DM masses $m_\chi = 59$\,keV
(dot-dashed) and 105\,keV (solid) for vector (blue) and
pseudo-scalar (green) operators. Since the pseudo-scalar
cross section is sensitive to the DM mass, the left axis
adopts a scale of $\mathcal O(1000)$ in blue for the vector
case while the right one of $\mathcal O(10)$ in green for
the pseudo-scalar one. ({\bf Right}) The atomic effects on
the thermally averaged total cross section with velocity
weight $\langle \sigma_{\chi e} v_\chi \rangle$ for
vector (blue) and pseudo-scalar (pseudo-scalar) cases.
Comparsion has been made between the scattering cross
section with a bound (thick) and free (thin) electron
while the bound/free ratios are shown in red color
according to the right axis scale.}
\label{fig:xsec}
\end{figure}

The left panel of \gfig{fig:xsec} shows the typical differential and
total cross sections for $m_\chi = (59, 105)$\,keV.
Since the scalar, vector, axial vector, and tensor
interactions share the same matrix element structure
and consequently the same differential spectrum,
only vector (blue) and pseudo-scalar (green) curves
are shown. For comparison, the electron recoil energy
from the DM absorption on a free electron at rest
takes a fixed value (gray),
$T_r = m^2_\chi / 2 (m_e + m_\chi)$ which is derived
without approximation. Although the spectrum
widens due to atomic effect, the peak is still at exactly
the same location as the free case. For $m_\chi = (59, 105)$\,keV,
the peak position is $T_r = (3.05, 8.95)$\,keV, respectively.
We have adjusted the $y$-axis scales for the vector (left blue)
and pseudo-scalar (right green) cases to make the $m_\chi = 105$\,keV
peaks with roughly the same height. Since the matrix element
$|\mathcal M|^2$ has $m_\chi - \Delta E_{nl}$ dependence for the
vector operator and $(m_\chi - \Delta E_{nl})^3$ for the
pseudo-scalar one, the $m_\chi = 59$\,keV peak heights scale
accordingly. With major contribution coming from the outer
shell electrons, the binding energy $E_b = (0.16, 0.012)$\,keV
for ($4p$, $5p$) shells can be negligibly small. Then the
peak height scales with $m_\chi - T_r$ and $(m_\chi - T_r)^3$,
respectively. For the vector (pseudo-scalar) case, the peak
height reduces by a factor of 0.58 (0.20). This explains why
the pseudo-scalar peak is only around $1/3$ of the vector
counterpart for $m_\chi = 59$\,keV. The $T_r$ dependence arises
from the $K$-factor and energy gain $\Delta E_{nl} = T_r - E_{nl}$.
Although the $m_\chi - T_r$ dependence is quite different,
the spectrum shape has only slight difference among operators.
This is because the two cases shown in the left panel of
\gfig{fig:xsec} has $m_\chi \gg T_r$. Consequently, the
$T_r$ spectrum is mainly determined by the $K$-factor which
is universally shared. This allows a unique probe that is
model-independent, or at least operator-independent, to some
extent. Once detected, fermionic DM absorption even allows
in-situ measurement of the atomic $K$-factor.

The right panel of \gfig{fig:xsec} shows the total cross section
$\langle \sigma v_\chi \rangle$ as a function of the DM mass.
With larger $m_\chi$, the cross section also becomes larger.
For light DM, $m_\chi \ll m_e$, \geqn{eq:sigmachie} indicates
that $\sigma^P_{\chi e} v_\chi$ for pseudo-scalar type scales
as $m^4_\chi$ while the others as $m^2_\chi$. The curves for
free electron scattering are fully consistent with the expected
scaling behaviors. Including atomic effects would reduce the
total cross section due to binding energy of the initial electron.
For inner shells, such as $E_b = (4.53, 0.95)$\,keV for ($2p$, $3p$)
electrons, the binding energy can be as large as the recoil
energy $T_r$ to significantly reduce the event rate. The green
lines show the ratio between the total cross sections with bound
and free electrons. A reduction of $0.5 \sim 0.7$ can happen.
With smaller $m_\chi$, the energy release is also smaller and
consequently harder to overcome the atomic bound energy which
leads to a larger suppression in the total cross section.

\subsection{Confronting the Xenon1T and PandaX-II Data}

In 2020, both Xenon1T and PandaX-II collaborations published their
electron recoil spectrum \cite{XENON:2020rca,PandaX-II:2020udv}.
An excess around $(2 \sim 3)$\,keV appears in the Xenon1T data with
significance reaching $3\sigma$. This could be explained by the
$\beta$ decays of tritium at $3.2 \sigma$ with a concentration
in xenon of $(6.2 \pm 2.0) \times 10^{-25}$\,mol/mol, but
``{\it such a trace amount can neither be confirmed nor
excluded}\,'' \cite{XENON:2020rca}. The PandaX-II data is
fully consistent with such a founding \cite{PandaX-II:2020udv}.
Since a sub-MeV fermionic DM absorption also leaves a sharp peak
at low recoil energy as shown in \gfig{fig:xsec}, confronting the
the Xenon1T and PandaX-II data can also provide a meaningful
constraint on the preferred parameter space.

The event rate of DM direct detection,
\begin{eqnarray}
  \frac {d N}{d T_r}
=
  N_T \frac {\rho_\chi}{m_\chi} t
\times
  \epsilon(T_r)
  \frac {d \langle \sigma_{\chi e} v_\chi \rangle}{d T_r},
\end{eqnarray}
scales with for the number of Xenon atoms $N_T$, DM local
number density $\rho_\chi / m_\chi$,
and run time $t$. The Xenon1T analysis uses 0.65 tonne-years
of data and PandaX-II 100 has tonne-days. In addition, the
detection efficiencies are basically constant above
$(3 \sim 4)$\,keV and decreases to 0 below there \cite{XENON:2020rca}.

\begin{figure}[t]
\centering
\includegraphics[width=0.495\textwidth]{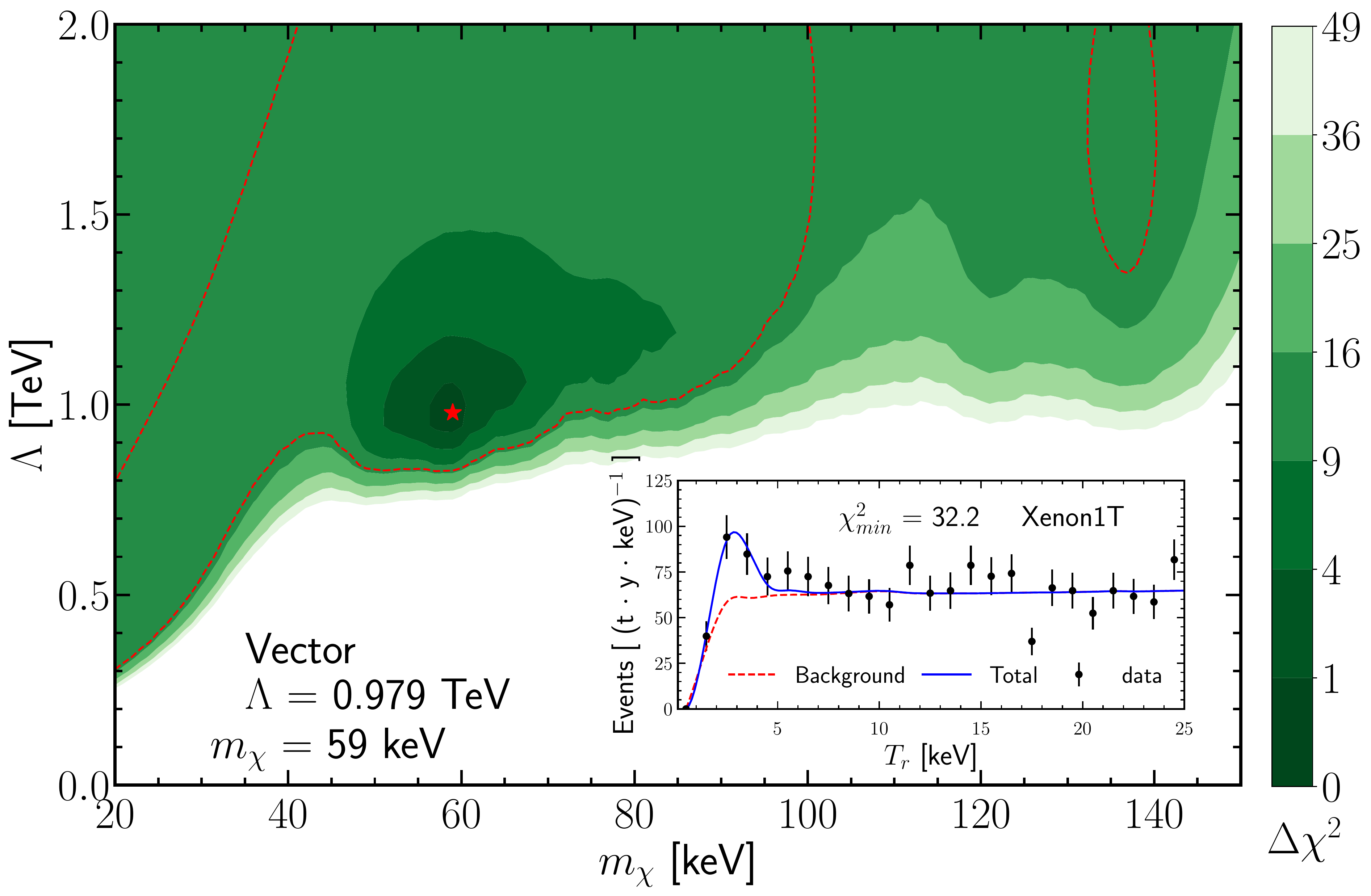}
\hfill
\includegraphics[width=0.48\textwidth]{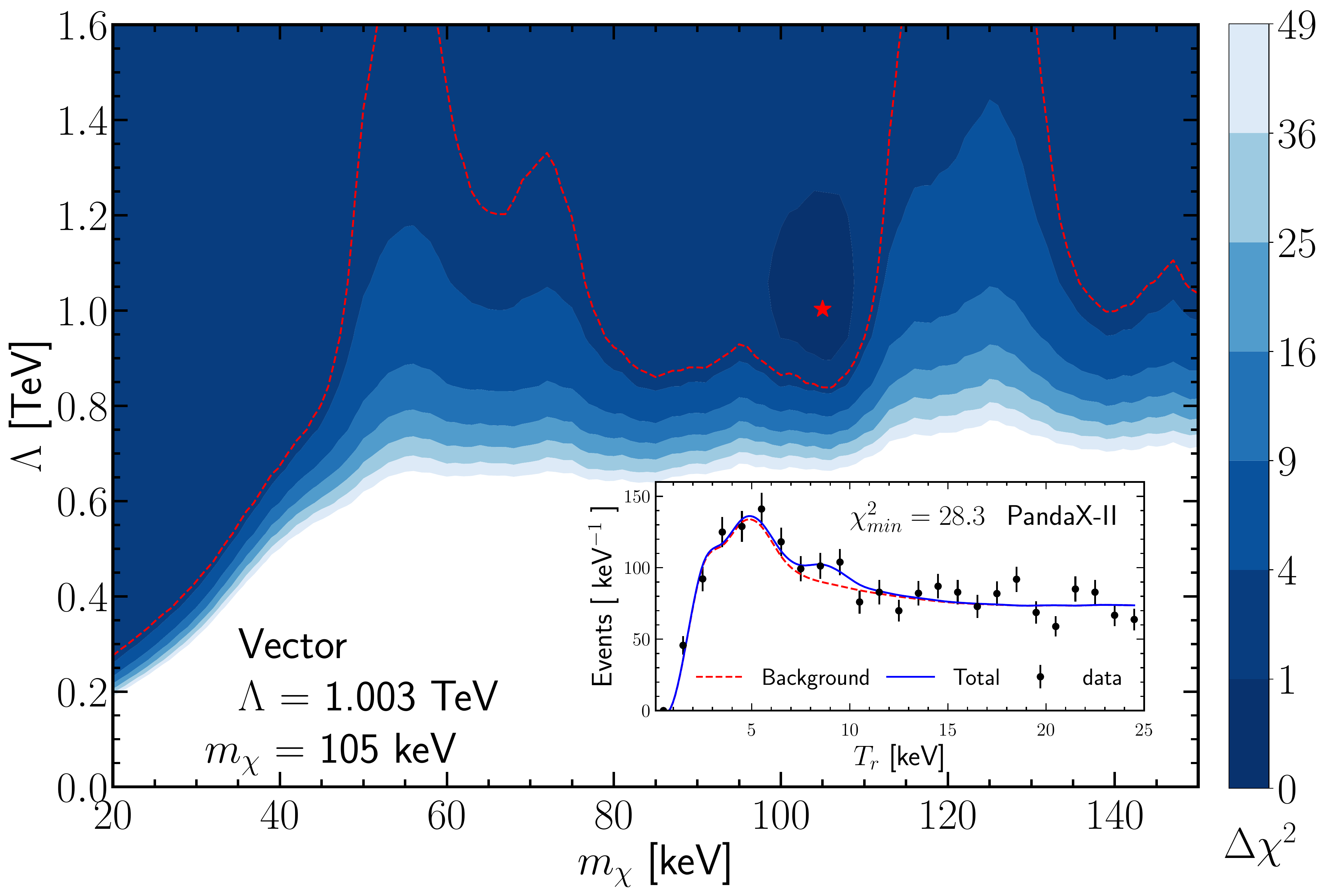}
\caption{The best fits (red star) and sensitivity contours
of fermionic DM absorption at Xenon1T ({\bf Left}) and
PandaX-II ({\bf Right}). Deeper color corresponds to smaller
$\Delta \chi^2$ and the white region has been excluded with
$\Delta \chi^2 > 49$.
For illustration, only the vector case is shown with best
fit values $m_\chi = 59$\,keV and $\Lambda = 0.979$\,TeV
at Xenon1T while $m_\chi = 105$\,keV and $\Lambda = 1.003$\,TeV
at PandaX-II. The red dashed contour shows the 95\% C.L. for
comparison with other plots. In addition to data points (black),
the inset plots demonstrate the background (red dashed) and
total (blue solid) event rates with the corresponding best
fit values.}
\label{fig:BothChi2}
\end{figure}

We adopt the analytical $\chi^2$ analysis
\cite{Ge:2012wj,Ge:2016zro}, whose advanced version is summarized
in \gapp{app:chi2fit}, to estimate the sensitivity.
In addition to the fermionic
DM absorption signal, background estimations are taken from
the experimental papers \cite{XENON:2020rca}. The results are shown in
\gfig{fig:BothChi2} for the fit with Xenon1T (left panel) and
PandaX-II (right panel) data. Since the different DM absorption
operators share roughly the same spectrum shape,
one representative vector
case can already show the features clearly. For Xenon1T, the
best fit is at $m_\chi = 59$\,keV and $\Lambda = 1$\,TeV,
being consistent with \cite{Dror:2020czw}. The inset plot
shows the signal and background curves with the best fit
values. The fermionic DM absorption signal
with peak at $T_r = 3.1$\,keV can fit the Xenon1T excess very nicely.
Comparing with the background-only fit, the $\chi^2_{\rm min}$
decreases from 46.3 to 32.2.
The decreasing edge for $m_\chi \rightarrow 20$\,keV is due
to two major reasons: 1) the cross section decreases with
$m^2_\chi$ in this region and 2) the efficiency further
suppresses its event rate to make it less sensitive.
Both needs compensation of a larger coupling strength,
or equivalently a lower cut-off scale $\Lambda$.
On the other side, the rising edge for
$m_\chi \rightarrow 150$\,keV at Xenon1T is due to the
abormally lower data point around $(17 \sim 18)$\,keV where
the recoil spectrum peaks. For PandaX-II data \cite{PandaX-II:2020udv},
the best fit is at $m_\chi = 105$\,keV
corresponding to the small excess at $T_r = (8 \sim 9)$\,keV.
Since the peak is not that significant, the $\chi^2_{\rm min}$
decreases by only less than 3 from 31.0 of the background-only
fit to 28.3.
Different from the Xenon1T data, the $T_r = (17 \sim 19)$\,keV
data points are higher than the expected background instead
which leads to a flat tail for $m_\chi \rightarrow 150$\,keV.

\section{The DM overproduction in the early universe}
\label{sec:overproduction}

A light DM is typically produced non-thermally. This is because the
thermal freeze-out give a relic density
$\rho_\chi \propto m^2_\chi / \langle \sigma v \rangle$ as ratio
between the DM mass and the thermally averaged cross section
\cite{Kolb:1990vq}.
To obtain the correct relic density, the interaction strength
between DM and SM particles should scale linearly with its mass.
Consequently, a light DM typically has a very small coupling
with SM particles. Then it is difficult for the light DM to reach
thermal equilibrium with the environmental plasma before
the thermal freeze-out \cite{Bernal:2017kxu}. Their production
is usually realized by the so-called freeze-in mechanism
\cite{Asaka:2005cn,Asaka:2006fs,Gopalakrishna:2006kr,Page:2007sh,
Hall:2009bx}.
No matter how the DM relic density is generated, it cannot exceed
the observed value, $\Omega_{\rm dm} h^2\approx 0.12$
\cite{Planck:2018vyg}.

\subsection{Boltzmann Equation and Its Solution}
\label{sec:Boltzmann}

For the fermionic DM absorption on electron target, it is
intrinsically connected to electron and positron as demonstrated
by the operators in \geqn{eq:operator}. The light DM can be
produced by the pair annihilation process $e^+ e^- \to \chi\nu$.
The Boltzmann equation governing the evolution of DM number
density $n_\chi$ is,
\begin{eqnarray}
{d n_\chi \over dt }+ 3 H n_\chi
& = & 
  \int d \Pi_\chi d \Pi_\nu d \Pi_{e^-} d \Pi_{e^+}
  (2 \pi)^4 \delta^{(4)}(p_{e^+}+p_{e^-}-p_\chi - p_\nu)
\nonumber
\\
& \times &
\left[ |{ \cal M}|_{e^+ e^- \to \nu\chi}^2
f_{e^+}f_{e^-}(1-f_\nu)(1-f_\chi)
-
|{ \cal M}|_{\nu\chi\to e^+ e^- }^2f_\nu f_\chi(1-f_{e^+} )(1-f_{e^-})
\right],
\label{eq:Beq0}
\end{eqnarray} 
where $H$ is the Hubble parameter. For particle $i$, 
$f_i$ is its phase space distribution function and
$d\Pi_i \equiv d^3p_i / 2 E_i (2\pi)^3$
is the phase space integration element.

With the freeze-in mechanism,
the DM density increases from 0. During the production process,
$f_\chi \ll f_{e^+}, f_{e^-}$. For simplicity, the second term
of \geqn{eq:Beq0} can be omitted \cite{Hall:2009bx} and the
Boltzmann equation then reduces to,
\begin{eqnarray}
{d n_\chi \over dt }+ 3 H n_\chi
& = & 
\langle v_{\rm M\o l}\sigma_{e^+ e^-} \rangle n_{e^+}^{\rm eq}n_{e^-}^{\rm eq}, 
\label{eq:Beq1}
\end{eqnarray} 
where $v_{\rm M \o l}$ is the M$\o$ller velocity of incoming
electron/positron pair \cite{Gondolo:1990dk} and $n^{\rm eq}_{e^\pm}$
are their number density at thermal equilibrium. We will come
back to provide a detailed justification of this simplification
later.

To solve the Boltzmann equation, we introduce the DM yield
$Y \equiv n_\chi / s(T)$ as the ratio of DM number density $n_\chi$
over the entropy density $s(T)$ as a function of temperature
$T$. At the epoch of DM production,
the universe is dominated by radiation. Both the Hubble parameter
$H$ and entropy density $s(T)$ can evolve with temperature,
\begin{eqnarray}
  H = 1.66\sqrt{g_{*}} {T^2 \over M_{\rm P}},
\qquad
  s(T) = g_{*s}{2\pi^2 \over 45}T^3,
\end{eqnarray}
where $M_P=1.22\times 10^{19}~\rm GeV$ is the Planck mass.
The relativistic degrees of freedom $g_{*}=g_*(T)$ and
$g_{*s}=g_{*s}(T)$ associated with the energy and entropy
densities, respectively, are taken from \cite{Husdal:2016haj}
while more detailed discussions can be found in
\cite{Saikawa:2018rcs,Saikawa:2020swg}.
Although the DM yield keeps increasing, the effective
degrees of freedom are mainly contributed by the SM particles.
Then in terms of yield $Y$, the Boltzmann equation \geqn{eq:Beq1}
becomes,  
\begin{eqnarray}
  {dY \over dT}
=
- {45M_{\rm P} \over 
2\pi^2 (1.66\sqrt{g_*})\tilde g_{*s} T^6 } \langle v_{\rm M\o l}\sigma_{e^+ e^- } \rangle
  n_{e^+}^{\rm eq} n_{e^-}^{\rm eq},
\quad \mbox{with} \quad 
\tilde g_{*s} \equiv g_{*s} \left( 1 +{T \over 3 g_{*s}} {dg_{*s} \over d T}\right)^{-1}.
\label{eq:Beq2}
\end{eqnarray}
The minus sign in $dY / dT$ arises because the temperature
decreases with time but the DM yield $Y$ increases.

The electron and positron annihilation $e^+ e^- \rightarrow \chi \nu$
happens when the temperature decreases to around the electron mass,
$T \sim 2 m_e$ \cite{Lehmann:2020lcv}. Then the inverse process
$\chi \nu \rightarrow e^+ e^-$ starts to decrease
and the $e^\pm$ density becomes exponentially suppressed.
In other words, the Fermi-Dirac distribution
of electron (positron) can be approximated by the Maxwell-Boltzmann
distribution $f_{e^\pm}^{\rm eq} \approx e^{- E/T}$.
For quantitative illustration, the typical electron (positron) energy
is $E_e \approx 2.27$\,MeV at $T \approx 1$\,MeV to
$E_e \approx 1.01$\,MeV at $T\approx 0.4$\,MeV. Correspondingly,
the Maxwell-Boltzmann distribution gives $e^{- E/T} = 0.104$ at
$T\approx 1$\,MeV and $0.080$ at $T\approx 0.4$\,MeV,
which are very close to the Fermi-Dirac values $0.094$
and $0.074$, respectively.  This further
simplifies the thermally averaged cross section
$\langle v_{\rm M\o l} \sigma_{e^+ e^-} \rangle$ 
\cite{Gondolo:1990dk},
\begin{eqnarray}
  \langle v_{\rm M \o l} \sigma_{e^+ e^-} \rangle n_{e^+}^{\rm eq} n_{e^-}^{\rm eq}
=
{T\over 8\pi^4} \int_{4m_e^2}^\infty ds (s-4m_e^2)\sqrt{s}K_1\left({\sqrt{s}\over T}\right)\sigma_{e^+ e^-}(s), 
\label{eq:vsigmaop}
\end{eqnarray}
where $K_1$ is the first modified Bessel function of second kind
and $s\equiv (p_{e^-} + p_{e^+})^2$ is the electron positron
invariant mass squared.

The solution of the Boltzmann equation in \geqn{eq:Beq2}
can be obtained by integrating the temperature $T$ from
neutrino decoupling $T_{\rm max} \approx 1$\,MeV, 
\begin{eqnarray}
  Y(T)
=
  {45 M_{\rm P} \over 16\pi^6}
  \int_{T}^{T_{\rm max}}  
{d \widetilde T \over (1.66\sqrt{g_*})\tilde g_{*s} \widetilde T^5 } 
\int_{4m_e^2}^\infty ds (s-4m_e^2)\sqrt{s}K_1\left({\sqrt{s}\over \widetilde T}\right)\sigma_{e^- e^+}(s).
\label{eq:Y0}
\end{eqnarray}
For the DM absorption operators in \geqn{eq:operator},
the $e^- e^+ \to \nu\chi$ cross section is a function
of the invariant mass $s$,
\begin{subequations}
\begin{eqnarray}
  \sigma^S_{e^+ e^-}
& = &
{1\over \Lambda^4}
 {\sqrt{s- 4m_e^2}(s-m_\chi^2)^2 \over 32\pi s\sqrt{s} },
\\
  \sigma^P_{e^+ e^-}
& = &
{1\over \Lambda^4}
 {(s-m_\chi^2)^2 \over 32\pi\sqrt{s}\sqrt{s- 4m_e^2}},
\\
  \sigma^V_{e^+ e^-}
& = &
{1\over \Lambda^4},
 {(s+2m_e^2)(2s+m_\chi^2)(s-m_\chi^2)^2 \over 48\pi s^2\sqrt{s}\sqrt{s- 4m_e^2}}
\\
  \sigma^A_{e^+ e^-}
& = &
{1\over \Lambda^4}
 {[2s(s-4m_e^2) +m_\chi^2(s+2m_e^2) ](s-m_\chi^2)^2 \over 48\pi s^2\sqrt{s}\sqrt{s- 4m_e^2}},
\\
  \sigma^T_{e^+ e^-}
& = &
{1\over \Lambda^4}
 {(s+2m_e^2)(s+2m_\chi^2)(s-m_\chi^2)^2 \over 12\pi s^2\sqrt{s}\sqrt{s- 4m_e^2}}.
\end{eqnarray}
\label{eq:sigmaop}
\end{subequations}

\begin{figure}[t]
\centering
\includegraphics[width=12cm]{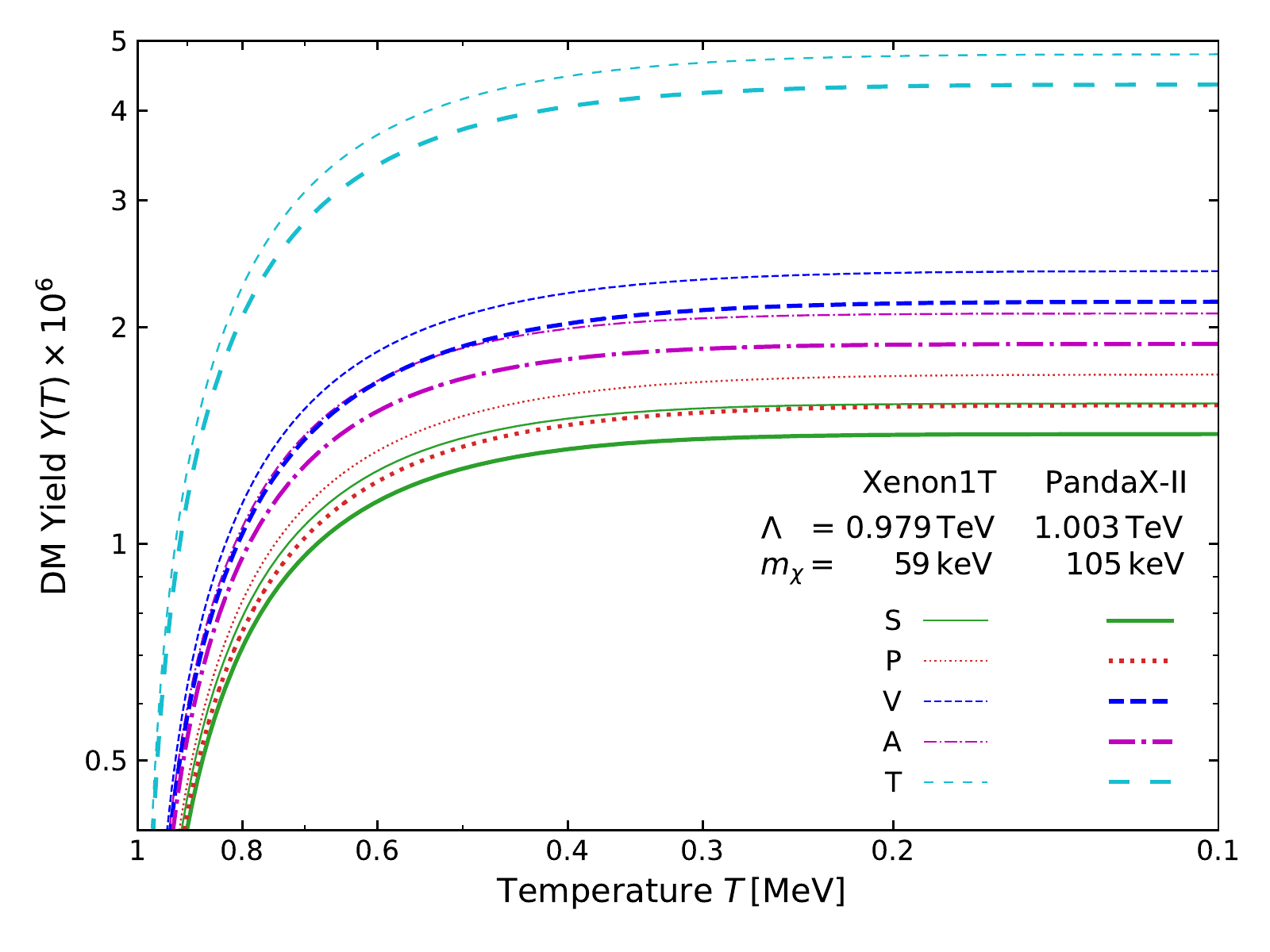}
\caption{The evolution of the DM yield $Y(T)$ as function of
temperature $T$ for the DM absorption operators with the best
fit values $m_\chi =59\,(105)$\,keV and $\Lambda=0.979\,(1.003)$\,TeV
of the Xenon1T (PandaX-II) data as shown in thin (thick) lines.
The different fermionic DM absorption operators are shown with
different line types and colors: scalar (S: green solid),
pseudo-scalar (P: red dotted), vector (V: blue dashed), axial
vector (A: purple dash-dotted), and tensor (T: cyan long dashed).}
\label{fig:DMyield}
\end{figure}

\gfig{fig:DMyield} shows the evolution of the DM yield $Y(T)$
as a function of temperature $T$.
For illustration, we adopt the scale $\Lambda=1$\,TeV and
DM mass $m_\chi=60$\,keV. For all the five DM absorption operators,
the DM yield converges when the Universe cools down to
$T\sim 0.4$\,MeV. In the light DM limit,
$m_\chi \ll m_e \lesssim \sqrt s$, the cross sections in \geqn{eq:sigmaop}
have quite simple scaling behaviors,
$\sigma^{S,P,V,A,T}_{e^+ e^-} \approx (\frac 1 8, \frac 1 8, \frac 1 6,
\frac 1 6, \frac 1 3) \times s / 4 \pi \Lambda^4$. There is no big difference
among scalar and pseudo-scalar operators or among the vector and axial vector
ones as correctly reflected in \gfig{fig:DMyield}. With larger cross section,
the DM yield converges to a larger value. Between the scalar/pseudoscalar group
and the vector/axial vector group, the converging values of the DM yield
roughly differs by a factor of 3/4 which is consistent with the relative
size among the cross sections,
$\sigma^{S,P}_{e^+e^-} / \sigma^{V,A}_{e^+e^-} \approx 3/4$. The small
deivation comes from the finite size of the DM mass $m_\chi = 60$\,keV
that is used in \gfig{eq:sigmaop}. 
The DM yield scales linearly with the $e^+ e^- \rightarrow \chi \nu$
cross section. We can also check that between the vector/axial vector
group and the tensor operator, the factor of 2 difference is also
consistent with both sides.

\subsection{Consistency Check of the Simplified Boltzmann Equation}
\label{sec:consistency}

Before proceeding to constrain the coupling strength of the
DM absorption operators, we need to first justify the omission
of the second term in \geqn{eq:Beq0} for consistency check.
The freeze-in production
of DM spans from neutrino decoupling ($T\sim 1$\,MeV) to the
end of $e^+e^-$ annihilation ($T\sim 0.1$\,MeV) 
\cite{Lehmann:2020lcv}. During this period of time,
electrons and positrons are still in equilibrium with
the thermal bath and therefore follow the
Fermi-Dirac distribution, $f_{e^\pm} \equiv {1/(e^{E_{e^\pm}/T}+1)}$.
The most likely value of the electron/positron energy
maximizes the energy distribution
$\sqrt{E_{e}^2 -m_e^2}E_{e} f_{e}^{\rm eq}$.
For example, the peak energy is
$E_e \approx 2.27$\,MeV at $T\approx1$\,MeV and
$E_e \approx 1.01$\,MeV at $T\approx 0.4$\,MeV.
Correspondingly, the phase space distribution function is
roughly $f_{e}^{\rm eq}\sim 0.094$ (0.074) at $T\approx 1\,\rm MeV$
($0.4\,\rm MeV$). Therefore, we can approximate
$(1-f_{e^\pm})\approx 1$ in the second term of \geqn{eq:Beq0}.

The cosmological evolution of DM density is related to the observed
number density $n^0_\chi = \rho_\chi / m_\chi$ today.
Especially, the maximal value of $n_\chi$ at the end
of freeze-in process is $n^0_\chi / a^3$ neglecting the
possible DM decay. At temperature
$T = 0.4$\,MeV, the scale factor $a \equiv 1/(1+z)$ or equivalently
the redshift is $z \approx 1/a \approx 1.6 \times 10^9$ \cite{Young:2016ala}.
The upper limit is reached when the freeze-in process
contributes the full DM density. Since the DM is not in
thermal equilibrium, one can only use the typical energy $\overline E_\chi$
to estimate the size of the phase space distribution function,
\begin{eqnarray}
  n_\chi
= \int {d^3p\over (2\pi)^3}f_\chi
\approx {1 \over 2\pi^2} f_\chi(\overline E_\chi) \overline E_\chi^3
\lesssim { n^0_\chi \over a^3}
\approx {\rho^0_\chi \over m_\chi} z^3
\approx 
4\times 10^{-5}~{\rm MeV^3}{ {\rm keV} \over m_\chi}.
\end{eqnarray}
The typical DM energy can be determined using the
energy conservation condition $E_\chi + E_\nu = E_{e^+} + E_{e^-}$
with the typical electron and positron energies $E_{e^\pm}$.
In a head-on collision and $E_{e+} = E_{e^-}$, the neutrino energy
$E_\nu = \sqrt{E_\chi^2 - m_\chi^2}$ is directly related to the
DM momentum and the DM energy is solved to be
$E_\chi =[m_\chi^2 + (E_{e^-}+E_{e^+})^2 ]/[2(E_{e^-}+E_{e^+})]$.
For a sub-MeV DM with $m_\chi \ll E_{e^\pm}$, the DM energy has
a lower limit,
$E_\chi \gtrsim (E_{e^-}+E_{e^+})/2$ which is approximately 1\,MeV
at $T\approx 0.4\,\rm MeV$. The typical DM phase space factor is
then bounded from above, 
\begin{eqnarray}
  f_\chi
\lesssim
  4 \times 10^{-5}~{\rm MeV^3}
  {{\rm keV} \over m_\chi}
  {2\pi^2 \over \overline E_\chi^3} 
\approx
  8\times 10^{-4} { {\rm keV} \over m_\chi},
\end{eqnarray}
which is truly small comparing the electron counterpat,
$f_e \sim 0.1$.

For the neutrino phase space distribution $f_\nu$,
there are two components. One is the standard cosmic
neutrino background that was in thermal equilibrium and
follows the Fermi-Dirac distribution
$f_\nu^{\rm eq}\equiv 1/(e^{E_\nu/T}+1)$.
The other one is the associated production from
$e^+ e^- \rightarrow \chi \nu$ that shares a similar
contribution ($f_\nu \lesssim 10^{-3}$) as DM.
So we can approximate $f_\nu \approx f_\nu^{\rm eq}$.
Based on a similar analysis as for $f_{e}^{\rm eq}$ by
requiring that $E_\nu^2f_\nu^{\rm eq}$ takes its maximal
value, one can obtain the typical neutrino energy
$E_\nu\approx 2.22\,(0.89)\,\rm MeV$ at $T\approx 1\,(0.4)\,\rm MeV$.
This further leads to a typical phase space distribution
$f_\nu^{\rm eq}\approx 0.1$. With massless neutrino,
this estimation is independent of $T$ since its phase
space distribution function only depends on the ratio $E_\nu/T$.
Therefore, we also approximate $1 - f_\nu \approx 1$
to very good accuracy.   

Putting things ($f_{e^\pm}$, $f_\chi$, and $f_\nu$) together,
we can justify the simplification of omitting the second
term in \geqn{eq:Beq0}. At the converging point ($T\approx 0.4\,\rm MeV$)
of DM yield, the phase space factor ratio 
\begin{eqnarray}
  {f_{e^+} f_{e^-} \over f_\nu \bar f_\chi}
\gtrsim
  61 {  m_\chi  \over {\rm keV} }
\gg 1, 
\end{eqnarray}
for the sub-MeV DM $1\,{\rm keV} \lesssim m_\chi \lesssim 1\,{\rm MeV}$
clearly indicates that the first term of \geqn{eq:Beq0} dominates
over the second one \cite{Hall:2009bx}.

\subsection{Constraints from DM Overproduction}
\label{sec:OverproductionConstraints}

To estimate the DM yield today $Y_0$, we assume there
is no other mechanisms to produce/deplete DM after its production
after the converging point $T_{\rm min}$. Then $Y_0=Y(T_{\rm min})$
with time-independence
since both the DM number density $n_\chi$ and the entropy density
$s$ scales as $1/a^3$. The DM relic density is estimated as,  
\begin{eqnarray}
  \Omega_\chi h^2 
= {2 m_\chi Y_0 s_0 h^2 \over \rho_c},
\end{eqnarray}
where $s_0=2970\,\rm cm^{-3}$ is the present entropy density and
$\rho_c = 1.054\times 10^{-5} h^2\,\rm GeV\,cm^{-3}$ the critical
density. The Hubble constant $h=0.67$ is in unit of
$100\,\rm km\,s^{-1}\,Mpc^{-1}$. Since both DM and its anti-particle
can be produced, there is a factor of 2 in the above estimation.
Requiring the DM relic density to be
less than the measured value, $\Omega_\chi h^2 \lesssim 0.12$, sets
a lower bound on the new physics scale $\Lambda$ of the DM
absorption operators in \geqn{eq:operator}. In other words, the DM
relic density cannot be overproduced.

\begin{figure}
\centering
\includegraphics[width=16cm]{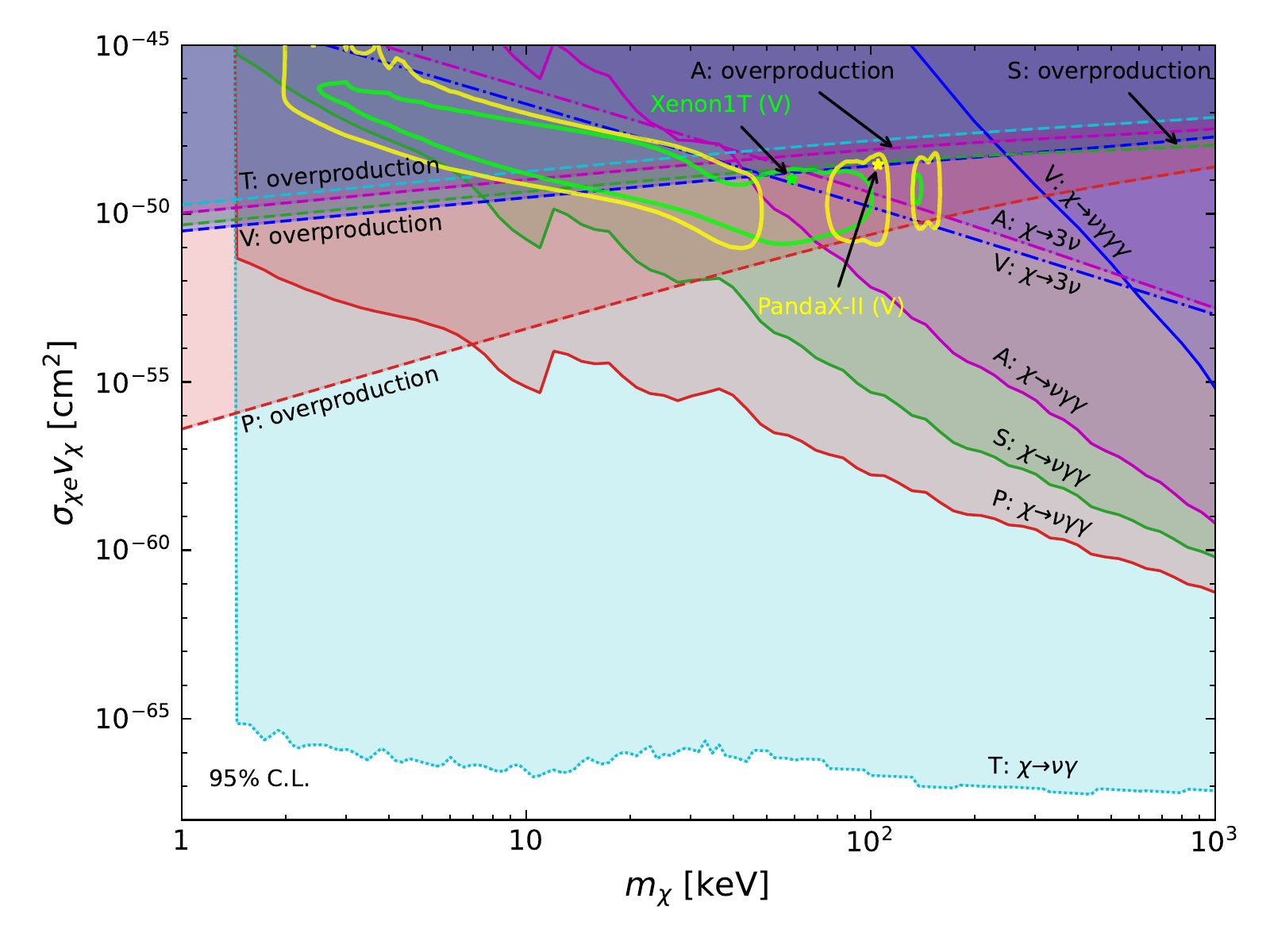}
\caption{The 95\% C.L. constraints on the fermionic absorption operators
from the DM overproduction (filled regions with dashed boundary),
the cosmological evolution history (filled regions for invisible decay
$\chi \rightarrow 3 \nu$ with dot-dashed boundary), as well as
the astrophysical X(gamma)-ray data (filled regions for visible decays
$\chi \rightarrow \nu \gamma (s)$). The exclusion regions are
filled with colors for scalar (S: green), pseduo-scalar (P: red),
vector (V: blue), axial vector (A: magenta), and tensor (T: cyan)
operators, respectively. Note that the exclusion region for the
tensor operator from astrophysical X(gamma)-ray constraint (T:
$\chi \rightarrow \nu \gamma$) uses dotted boundary to indicate that
this constraint is subject to uncertainty in the regularization
scheme. For comparison, the best fit points (star) and 95\% C.L.
contours (contour) for the Xenon1T (light green) and PandaX-II
(yellow) are also shown with the vector-type operator as an example.}
\label{fig:constraints}
\end{figure}

\gfig{fig:constraints} shows the constraints on the direct
detection cross section $\sigma_{\chi e}v_\chi$. The excluded
parameter space by the DM overproduction is shown as filled region
with dashed boundary. These DM overproduction
bounds are quite stringent in the low mass region, especially
for the pseudo-scalar case. This is because for $m_\chi \ll 2m_e$,
the cross-section scales as $\sigma_{e^- e^+} \propto 1/\Lambda^4$
and becomes independent of the DM mass according to \geqn{eq:sigmaop}.
Consequently, the DM yield $Y$ estimated by \geqn{eq:Y0} is not
sensitive to $m_\chi$ and the DM relic density scales as
$\Omega_\chi h^2 \propto m_\chi/\Lambda^4$. However, the direct
detection cross section in \geqn{eq:sigmachie}
scales as power of the DM mass, $\sigma_{\chi e}^{P}v_\chi
\propto m^4_\chi / \Lambda^4 \propto m_\chi^3 \Omega_\chi$
for the pseudo-scalar case and $\sigma_{\chi e}^{V,S,A,T}v_\chi \propto
m_\chi^2/\Lambda^4 \propto m_\chi \Omega_\chi$ for the others.
This explains why the pseudo-scalar bound decreases faster with $m_\chi$.
For the scalar, (axial-)vector, and tensor cases, they have a similar
sensitivity around $10^{-50}\,\rm cm^2$ to $10^{-47}\,\rm cm^2$ for the
DM mass from 1\,keV to 1\,MeV. 

Using the more exactly calculated cross section
$\sigma_{\chi e} v_\chi$ as summarized in \geqn{eq:sigmachie},
instead of the approximation $m^2_\chi / 4 \pi \Lambda^4$ in
the $m_\chi \ll m_e$ limit, for the vertical axis has some
advantage. As explained below \geqn{eq:sigmaop}, the
$e^+ e^- \rightarrow \chi \nu$ cross section almost degenerates
between scalar ($\sigma^S_{e^+ e^-}$) and
pseudo-scalar ($\sigma^P_{e^+ e^-}$) operators. So one may
expect their overproduction limits not to differ much and
hard to distinguish in \gfig{fig:constraints} is the universal
$m^2_\chi / 4 \pi \Lambda^4$ is adopted. In contrast, the
direct detection cross sections $\sigma^S_{\chi e} v_\chi$
and $\sigma^P_{\chi e} v_\chi$ have quite different scaling
behaviors as discussed around \geqn{eq:sigmachie}. So we
can still see the clear difference between the scalar and
pseudo-scalar cases in \gfig{fig:constraints}. In addition,
$\sigma_{\chi e}$ also has the advantage of being able to
reflect the realistic direct detection signal strength.

For comparison, the best fit points (star) with the Xenon1T
(light green) and PandaX-II (yellow) are also shown. The best fit
values are taken from \gfig{fig:BothChi2} for the vector case.
We can see
that for both data sets, the best fit points are already at the
boundary of the DM overproduction constraints.
Also shown are the 95\% C.L. allowed regions for the Xenon1T
(light green contour) and PandaX-II (yellow contour) data.
Consistent with the 95\% C.L. contours in \gfig{fig:BothChi2},
the 95\% C.L. allowed parameter space here is divided into
2 (3) regions for the Xenon1T (PandaX-II) data. The regions
extend significantly down to a few keV of the DM mass $m_\chi$.
More discussions about the cosmological constraint from
the Universe expansion history and the astrophysical ones from
various X(gamma)-ray observations will be discussed later in
\gsec{sec:cosmo} and \gsec{sec:astro}, respectively.

\section{DM decay}
\label{sec:DMDC}

One important feature of the DM absorption process is that only one
DM particle can appear in the process as demonstrated by the general
fermionic DM absorption operators in \geqn{eq:operator}. A natural
consequence is that DM is unstable and can decay into light SM particles
unless forbidden by kinematics. Since the DM particle is neutral,
electron and positron should appear as a pair if such decay topology
is possible. For $\chi \rightarrow e^+ e^- + \cdots$ to happen,
the DM mass has to be larger that twice of the electron mass,
$m_\chi > 2 m_e$, which is already outside the mass range
considered in the current paper. The only possible decay products
are the neutrinos and photons.

\begin{figure}[t]
\begin{tikzpicture} [mystyle]
   \begin{scope}[shift={(0,1)}]
    \draw[f] (0, 0) node[left]{$ \chi $} -- (2,0);
    \draw[f] (2, 0) -- (4, 0)node[right] {$ \nu$};
\draw [fr={0.6}{ 0 }] (2,0) to [out=-180, in=180] (2,-1.5);
\draw [fr={0.4}{-180}] (2,0) to [out=0, in=0] (2,-1.5);
\node at (2.8, -0.7){$ e$};
\node at (1.3, -0.7){$e$};
\draw[v] (2, -1.5) -- (2,-3) node [below, xshift =10 pt,yshift =10 pt ] {$\gamma$};
 \draw[circlestyle] (2,0) circle (0.2);
  \node at (2, -3.5){(a)};
  \end{scope}
\end{tikzpicture}
\quad
\begin{tikzpicture} [mystyle]
   \begin{scope}[shift={(0,1)}]
    \draw[f] (0,0) node[left]{$ \chi$} -- (2,0);
    \draw[f] (2, 0) -- (4,0)node[right] {$ \nu$};
    \draw[f] (2, 0) -- (1,-1.5)node[midway, left, xshift =-5pt,yshift =5 pt] {$ e$};
    \draw[fb] (2, 0) -- (3,-1.5)node[midway, right, xshift =5pt,yshift = 5 pt] {$ e$};
     \draw[f] (1, -1.5) -- (3,-1.5)node[midway, below, yshift =-5pt] {$ e $};
   \draw[v] (3, -1.5) -- (3,-3) node[below, xshift =10 pt,yshift =10 pt ]{$ \gamma$};
    \draw[v] (1, -1.5) -- (1,-3) node[below, xshift =10 pt,yshift =10 pt ]{$ \gamma$};
   \draw[circlestyle] (2,0) circle (0.2);
    \node at (2,-3.5){(b)};
   \end{scope}
\end{tikzpicture}
\quad
\begin{tikzpicture} [mystyle]
   \begin{scope}[shift={(0,1)}]
    \draw[f] (0,0) node[left]{$ \chi$} -- (2,0);
    \draw[f] (2, 0) -- (4,0)node[right] {$ \nu$};
    \draw[f] (2, 0) -- (0.5,-1.5)node[midway, left, xshift =-3pt,yshift =5 pt] {$e$};
    \draw[fb] (2, 0) -- (3.5,-1.5)node[midway, right, xshift =5pt,yshift = 5 pt] {$e$};
     \draw[f] (0.5, -1.5) -- (2,-1.5)node[midway, below, yshift =-5pt] {$e $};
     \draw[f] (2, -1.5) -- (3.5,-1.5)node[midway, below, yshift =-5pt] {$e$};
   \draw[v] (3.5, -1.5) -- (3.5,-3) node[below, xshift =10 pt,yshift =10 pt ]{$ \gamma$};
    \draw[v] (2, -1.5) -- (2,-3) node[below, xshift =10 pt,yshift =10 pt ]{$ \gamma$};
    \draw[v] (0.5, -1.5) -- (0.5,-3) node[below, xshift =10 pt,yshift =10 pt ]{$ \gamma $};
   \draw[circlestyle] (2,0) circle (0.2);
    \node at (2, -3.5){(c)};
   \end{scope}
\end{tikzpicture}
\quad
\begin{tikzpicture} [mystyle]
   \begin{scope}[shift={(0,1)}]
    \draw[f] (0,0) node[left]{$ \chi$} -- (2,0);
    \draw[f] (2, 0) -- (4,0)node[right] {$ \nu$};
    \draw[f] (2, 0) -- (1,-1.5)node[midway, left, xshift =-5pt,yshift =2pt ] {$ e$};
    \draw[fb] (2, 0) -- (3,-1.5)node[midway, right, xshift =5pt,yshift = 2pt ] {$ e$};
   \draw[fb] (3, -1.5) -- (3,-3) node[below, xshift =10 pt,yshift =10 pt ]{$ \bar \nu$};
    \draw[f] (1, -1.5) -- (1,-3) node[below, xshift =10 pt,yshift =10 pt ]{$ \nu$};
    \draw[v] (1, -1.5) -- (3,-1.5)node[midway, below, yshift =-5pt] {$ W$};
   \draw[circlestyle] (2,0) circle (0.2);
       \node at (2,-3.5){(d)};
   \end{scope}
\end{tikzpicture}
\quad
\begin{tikzpicture} [mystyle]
   \begin{scope}[shift={(0,1)}]
    \draw[f] (0, 0) node[left]{$ \chi $} -- (2,0);
    \draw[f] (2, 0) -- (4, 0)node[right] {$ \nu$};
\draw [fr={0.6}{ 0 }] (2,0) to [out=-180, in=180] (2,-1.2);
\draw [fr={0.4}{-180}] (2,0) to [out=0, in=0] (2,-1.2);
\node at (2.8, -0.7){$ e$};
\node at (1.2, -0.7){$ e$};
\draw[fb] (2, -2) -- (3,-3) node[below, xshift =10 pt,yshift =10 pt ]{$ \bar \nu$};
\draw[f] (2, -2) -- (1,-3) node[below, xshift =-10 pt,yshift =10 pt ]{$ \nu$};
\draw[v] (2, -1.2) -- (2,-2)node[midway, xshift = 8pt, yshift = 8pt, below] {$ Z$};
 \draw[circlestyle] (2,0) circle (0.2);
     \node at (2,-3.5){(e)};
  \end{scope}
\end{tikzpicture}
\caption{The representative Feynman diagrams contributing
to the DM visible decays $\chi \to \nu +\gamma$(s)
as well as the invisible decay $\chi \to 3\nu$. The blue
vertex is the DM absorption operator while the others are
SM interactions.}
\label{fig:chiDC}
\end{figure}
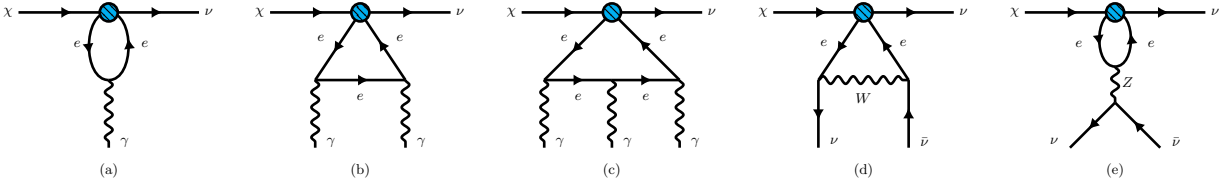

With the DM particle being a fermion, the final state has
to contain an odd number of neutrinos, including the visible
decay modes ($\chi \to \nu + \gamma$, $\chi \to \nu + \gamma\gamma$,
and $\chi \to \nu + \gamma\gamma\gamma$) as well as the
invisible mode $\chi \to 3 \nu$. All these can happen only
at loop level since the absorption operators in
\geqn{eq:operator} contains two electron fields. The leading
1-loop Feynman diagrams are listed in \gfig{fig:chiDC}.
For the visible decay modes, only electromagnetic interaction
of the SM is needed in addition to the absorption operator.
In contrast, the invisible decay mode $\chi \to 3 \nu $ requires
the SM weak interactions with $W/Z$ mediator.

To have a better understanding of the connection between
operators and decay processes, \gtab{tab:decaysummary}
summarizes the leading decay channels for each operator
highlighted with a checkmark ($\checkmark$). The cross
($\times$) indicates the decay channel that cannot be
generated at 1-loop level.
Although some can appear at two-loop level, they are
hugely suppressed by loop factor, the weak scale,
and/or phase space. For the $\chi\to \nu\gamma\gamma\gamma$
and $\chi\to 3\nu$ channels from the tensor operator,
the $\times !$ symbol is used to indicate that such
processes can be generated at 1-loop order but heavily
suppressed. In the following subsections we detail
our calculation for each operator.
The calculated decay width and spectrum are further used in
\gsec{sec:constraints} to derive the cosmological and
astrophysical constraints.

\begin{table}[h!]
    \centering
    \begin{tabular}{|c | c | c | c || c |}
        \hline
        \backslashbox{Operator}{Process} &
       $\chi\to\nu\gamma$    &
        $\chi\to\nu\gamma\gamma$    &
        $\chi\to\nu\gamma\gamma\gamma$   &
        $\chi\to 3 \nu$
          \\
        \hline 
         S: ${\cal O}_{e\nu\chi}^{S}$   & $\times$  &
        \checkmark  & $\times$ &  $\times$ \\
         \hline
         P: ${\cal O}_{e\nu\chi}^{P}$   & $\times$  &
        \checkmark  & $\times$  &  $\times$ \\ 
         \hline
       V: ${\cal O}_{e\nu\chi}^{V}$   & $\times$  & $\times$  &
        \checkmark  &  \checkmark \\ 
        \hline 
        A: ${\cal O}_{e\nu\chi}^{A}$   & $\times$  &
        \checkmark  & $\times$ & \checkmark \\
        \hline
        T: ${\cal O}_{e\nu\chi}^T$  &
       \checkmark & $\times$ &  $\times !$ &  $\times !$ \\
        \hline
    \end{tabular}
\caption{Contributions of the Fermionic DM absorption operators to
the visible ($\chi \rightarrow \nu \gamma (s)$ and invisible
($\chi \rightarrow 3 \nu$) decay channels. The one allowed
at one-loop level is shown with a checkmark (\checkmark) or a cross
($\times$) if otherwise. For those that allowed at one-loop level
but highly suppssed, an exclaimed cross ($\times !$) is used.}
\label{tab:decaysummary}
\end{table}

\subsubsection{The Scalar and Pseudoscalar Operators ${\cal O}_{e \nu \chi}^{S(P)}$}

For the operators ${\cal O}_{e \nu \chi}^{S,P}$ with an electron
scalar or pseudo-scalar current, the dominant decay mode is
$\chi \to \nu +\gamma\gamma$. Aa shown in \gfig{fig:chiDC}(b),
this process is generated through electron loop with two photons
attached to the electron loop. The amplitudes from loop
calculation are free from the UV divergence. Consequently,
the decay widths can be calculated exactly,
\begin{subequations}
\begin{eqnarray}
\Gamma_{\chi\to \nu\gamma\gamma}^S 
& = &  {1 \over \Lambda^4} {\alpha^2\over 2048\pi^5  m_e^2 m_\chi^3}
\int_0^{m_\chi^2} ds_{12}
s_{12}^2(m_\chi^2 - s_{12})^2
|F_S(\eta)|^2
,
\\
\Gamma_{\chi\to \nu\gamma\gamma}^P
& = &   {1 \over \Lambda^4} {\alpha^2\over 2048\pi^5  m_e^2 m_\chi^3}
\int_0^{m_\chi^2} ds_{12}
s_{12}^2(m_\chi^2 - s_{12})^2
|F_P(\eta)|^2,
\end{eqnarray}
\end{subequations}
where $\alpha \equiv e^4 / 4 \pi$ is the electromagnetic fine structure
constant. The integration variable $s_{12}$ is the squared
invariant mass of the two final-state photons, staring from 0 to
the maximal value $m^2_\chi$. The loop
functions $F_{S,P}(\eta)$ with $\eta\equiv s_{12}/m_e^2$ are,
\begin{subequations}
\begin{eqnarray}
 F_S(\eta) &\equiv& 
 {4\over \eta}-{\eta-4\over\eta^2}\ln^2{\sqrt{\eta-4}-\sqrt{\eta}\over\sqrt{\eta-4}+\sqrt{\eta}},
\\
 F_P(\eta) &\equiv&  -{1\over \eta}\ln^2{\sqrt{\eta-4}-\sqrt{\eta}\over\sqrt{\eta-4}+\sqrt{\eta}}. 
\end{eqnarray} 
\end{subequations}
In the limit of $m_\chi \ll 2m_e$ and consequently
$s_{12} \ll m^2_e$, the loop functions reduce to
$F_S(\eta) \approx 2/3$ and $F_P(\eta) \approx 1$.
Then the decay widths above can be approximated as,
\begin{subequations}
\begin{eqnarray}
\Gamma_{\chi\to \nu\gamma\gamma}^S 
& \approx &   9.4\times 10^{-20}{\rm sec^{-1}}
\left(m_\chi \over 200\,\rm keV\right)^7
\left(\rm TeV \over \Lambda\right)^4
,
\\
\Gamma_{\chi\to \nu\gamma\gamma}^P
& \approx &   
2.1\times 10^{-19}{\rm sec^{-1}}
\left(m_\chi \over 200\,\rm keV\right)^7
\left(\rm TeV \over \Lambda\right)^4.
\end{eqnarray}
\label{eq:chi2v2gapp}
\end{subequations}

For the 2-body mode $\chi \to \nu \gamma$ in \gfig{fig:chiDC}(a)
and the 4-body one $\chi\to \nu \gamma\gamma\gamma$ in
\gfig{fig:chiDC}(c), the electron loop contribution vanishes
due to the QED (quantum electrodynamics) charge conjugation
symmetry. This is because the involved electron currents
$\bar e e$ and $\bar e \gamma_5 e$ have an even parity
under charge conjugation transformation while the photon
field is odd. With an odd number of photons, the one-loop diagram
is odd and should vanish. Non-vanishing contribution can
only be generated at the 2-loop level involving both
QED and weak interactions. However, such a contribution
is severely suppressed by the loop factor and weak
scale such as $m_e^2/m_W^2$. Therefore, we can neglect the
single- and triple-photon contributions.  

The 3-body invisible decay $\chi \to 3\nu$ in the
last two diagrams of \gfig{fig:chiDC}
is also vanishing. The connection between the $\chi \nu$
fermion lines at the top and the other two neutrinos is
established through the $\mathcal O^S_{e \nu \chi}$ and
$\mathcal O^P_{e \nu \chi}$ operators. Correspondingly,
the effective current for the other two neutrinos after
loop integration should also be of the same scalar feature as
the $\bar e e$ and $\bar e \gamma_5 e$ counterpart in
the operator. Since the SM neutrinos are massless and
purely left-handed, the only possibility is
$\partial_\mu (\bar\nu_L \gamma^\mu \nu_L)$ with Lorentz
indices fully contracted. Then, the equation of
motion for a massless neutrino renders
$\partial_\mu \gamma^\mu \nu_L$ to vanish.
Although constructing a Majorana
mass term with only left-handed neutrinos is possible,
lepton number is not violated in weak interactions and
hence cannot appear without involving other new physics.

\subsubsection{The Vector Operator ${\cal O}_{e\nu\chi}^{V}$}

For the vector operator, the 2-body $\chi\to\nu\gamma$ and 3-body
$\chi \to \nu\gamma\gamma$ channels cannot arise at 1-loop level.
The vanishing of $\chi\to\nu\gamma$ is due to the gauge symmetry.
As mentioned in the previous paragraph, the electron current
$\bar e \gamma_\mu e$ in the vector operator ${\cal O}_{e\nu\chi}^{V}$
contributes to the loop mediator and needs to be integrated way
with a remaining photon field as external state. This feature
is generally parametrized as matrix element
$\langle\gamma(q,\epsilon)| \bar e \gamma_\mu e |0\rangle$.
In the presence of an external photon, the matrix element is
a linear function of its polarization vector $\epsilon^*_\mu$.
Another quantity that can provide a Lorentz index is the momentum
transfer $q_\mu$. The full matrix element also contains
a piece $\bar u_\nu \gamma^\mu P_L u_\chi$ from the
neutrino side,
\begin{eqnarray}
  \mathcal M
\equiv 
\left[
  A(q^2) q_\mu \epsilon^* \cdot q
+ B(q^2) \epsilon_\mu^*
\right]
  (\bar u_\nu \gamma^\mu P_Lu_\chi)
\equiv
  \epsilon_\mu^* {\cal M}^\mu.
\end{eqnarray}
The coefficients $A(q^2)$ and $B(q^2)$ correlate with
each other by the QED Ward identity: $q_\mu {\cal M}^\mu=0$
from replacing the photon polarization vector $\epsilon^*_\mu$
with its four-momentum $q_\mu$. Namely,
\begin{eqnarray}
\left[ A(q^2) q^2 + B(q^2) \right]
  q_\mu (\bar u_\nu \gamma^\mu P_L u_\chi)
= 0.
\end{eqnarray}
Since $q^\mu \equiv p_\chi^\mu - p_\nu^\mu$, the second term
$q_\mu (\bar u_\nu \gamma^\mu P_Lu_\chi)= \bar u_\nu(\slashed{p}_\chi -\slashed{p}_\nu)P_Lu_\chi=m_\chi \bar\nu_\nu P_Ru_\chi\neq 0$
is nonzero. Then the only solution is $B(q^2)=-A(q^2)q^2$ and
the effective current,
\begin{eqnarray}
  \langle\gamma(q,\epsilon)| \bar e\gamma_\mu e |0\rangle
=
  A(q^2)(q^2\epsilon^*_\mu - q_\mu \epsilon^* \cdot q),
\label{eq:Wardid}
\end{eqnarray}
vanishes due to the photon on-shell ($q^2=0$) and
transverse ($\epsilon^* \cdot q=0$) conditions.
For the 2-photon decay $\chi \rightarrow \nu \gamma \gamma$,
the loop part again contains 3 currents and hence vanishes
due to the QED charge conjugation symmetry.

The dominant visible decay channel is the 4-body process 
$\chi\to\nu\gamma\gamma\gamma$ whose matrix elment can be
generally parametrized as,
\begin{eqnarray}
{\cal M}_{\nu\gamma\gamma\gamma}  =   - {e^3 \over 16\pi^2\Lambda^2}
(\bar u_\nu \gamma_\mu P_L u_\chi) 
\epsilon_{1\alpha}^*\epsilon_{2\beta}^*\epsilon_{3\rho}^*\Pi^{\mu\alpha\beta\rho},
\label{eq:Achi2v3g}
\end{eqnarray}
where $\epsilon_i^*$ is the polarization vector of the $i$-th
final-state photon. The tensor $\Pi^{\mu\alpha\beta\rho}$ is
the reduced matrix element from the the electron loop in
\gfig{fig:chiDC}(c). Correspondingly, the spin averaged squared
matrix amplitude becomes, 
\begin{eqnarray}
\overline{|{\cal M}_{\chi\to\nu\gamma\gamma\gamma}|^2} 
=
 {16\alpha^3 \over \pi \Lambda^4}
  \left\{ - {1\over 4}
 \tr\left[\slashed{p}_\nu \gamma_\mu \slashed{p}_\chi\gamma_\nu\right] 
\left( {1 \over 8} \Pi^{\mu\alpha\beta\rho}\right)
\left( {1 \over 8}\Pi^{\nu}_{~\alpha\beta\rho}\right)
 \right\}. 
 \label{eq:A2chi2v3g}
\end{eqnarray}
The reduced matrix element $\Pi^{\mu\alpha\beta\rho}$
is evaluated first analytically by
Package-X \cite{Patel:2015tea} and then numerically by
COLLIER \cite{Denner:2016kdg}. The decay width is an
integral over the 4-body phase space $d \Phi_4$, 
\begin{eqnarray}
  \Gamma_{\chi\to\nu\gamma\gamma\gamma}^V
=
  {1 \over 3 !}  {1 \over 2m_\chi}\int d\Phi_4 \overline{|{\cal M}_{\chi\to\nu\gamma\gamma\gamma}|^2},
\label{eq:chi2v3gE}
\end{eqnarray}
with the factor $1/3!$ to avoid phase space overcounting
for the 3 identical photons.
In the limit of $m_\chi \ll 2m_e$, the decay width also has
a simple scaling behavior, 
\begin{eqnarray}
\Gamma_{\chi\to\nu\gamma\gamma\gamma}^V 
  \approx  2.6\times 10^{-29} {\rm sec^{-1}} 
\left(m_\chi \over 200\,\rm keV \right)^{13} \left( \rm TeV \over \Lambda \right)^4.
\label{eq:chi2v3gapp}
\end{eqnarray}

For the DM invisible decay $\chi \to 3\nu $, the loop diagrams
(d) and (e) in \gfig{fig:chiDC} suffer from UV divergence.
To make a reasonable estimation, we use the dimensional regularization
(DR) to tackle this issue. For the $W$-loop contribution,
the amplitude in the unitary gauge is
\begin{eqnarray}
  {\cal M}_{\chi\to 3 \nu}^W
=
- {\bar u_\nu\gamma^\mu P_L u_\chi \over \Lambda^2}
  {3 g_2^2 \over 64\pi^2}
\left[ 1 + {m_e^2 \over m_W^2 }
\left( {1 \over \epsilon} +{11 \over 6} +\ln{\Lambda^2 \over m_W^2} \right)
+\cdots \right]
  \left[ \bar u_{\nu_e}(p_1) \gamma_\mu P_L v_{\nu_e}(p_2) \right],
\end{eqnarray}
where $g_2$ is the weak coupling constant while $\mu$ the
dimensional regularization scale. The terms in the bracket is
from the $W$-loop for generating the electron neutrino pair,
and $\cdots$ stand for terms suppressed by higher powers of
$q^2/m_W^2$. Dropping the divergent piece $1/\epsilon$ altogether
with the terms proportional to $q^2/m_W^2$, the finite
amplitude used in our estimation is,
\begin{eqnarray}
{\cal M}_{\chi \to 3 \nu}^W \approx
-{1\over \Lambda^2}
{3g_2^2 \over 64\pi^2}
(\bar u_\nu\gamma^\mu P_L u_\chi)
[\bar u_{\nu_e}(p_1) \gamma_\mu P_L v_{\nu_e}(p_2)].
\end{eqnarray}
The amplitude and decay width depend on the neutrino flavor
that appears in the DM absorption operator. For
$(\bar \nu_e\gamma^\mu \chi_L)$, there are two diagrams by
exchanging the two electron neutrinos but only a single one
for $(\bar \nu_{\mu, \tau} \gamma^\mu \chi_L)$.
Considering this difference, the decay width for muon/tau
flavor is, 
\begin{eqnarray}
\Gamma_{\chi \to 3\nu}^{V}
\approx
{m_\chi^5 \over 1536\pi^3}
\left({3 g_2^2 \over 64\pi^2} \right)^2
{1 \over \Lambda^4}
\approx
3.66 \times 10^{-17}{\rm sec}^{-1} 
\left( m_\chi \over\rm 200~{\rm keV} \right)^5
\left( {\rm TeV} \over {\rm \Lambda} \right)^4,
\label{eq:chi23vV}
\end{eqnarray}
and for the electron neutrino in ${\cal O}_{e\nu\chi}^V$,
there is an additional enhancement factor 2. Note that the
$Z$ boson mediated diagram of \gfig{fig:chiDC}(e) is always
suppressed by $1/m^2_Z$ and hence can be neglected.

\subsubsection{The Axial-Vector Operator ${\cal O}_{e\nu\chi}^{A}$}

The dominant decay modes for the axial-vector operator
${\cal O}_{e\nu\chi}^{A}$ are the 3-body decay
$\chi \to \nu\gamma\gamma$ and $\chi \to 3\nu$.
This is because the 2-body mode $\chi\to \nu \gamma$
and the 4-body one $\chi\to \nu\gamma\gamma\gamma$
with an odd number of photons in the final state are
forbidden by the QED charge conjugation symmetry. 

First, the 3-body process $\chi \to \nu\gamma\gamma$ is
generated through the  similar electron loop as the
scalar and pseudo-scalar cases. The exact decay width is,
\begin{eqnarray}
  \Gamma_{\chi\to \nu\gamma\gamma}^A
= 
  {1 \over \Lambda^4} {\alpha^2\over 512\pi^5  m_e^4 m_\chi}
  \int_0^{m_\chi^2} ds_{12}
  s_{12}^2(m_\chi^2 - s_{12})^2
  |F_A(\eta)|^2,
\end{eqnarray}
where $s_{12}$ is again the squared invariant mass
of the two final-state photons and $\eta \equiv s_{12}/m_e^2$.
The new loop function $F_A(\eta)$,
\begin{eqnarray}
  F_A(\eta)
\equiv
- {1\over \eta}
- {1\over \eta^2}\ln^2{\sqrt{\eta-4}-\sqrt{\eta}\over\sqrt{\eta-4}+\sqrt{\eta}},
\end{eqnarray}
reduces to $F_A(\eta) \approx 1 / 12$
in the limit of $m_\chi \ll 2m_e$. Accordingly,
the decay width scales as 
\begin{eqnarray}
  \Gamma_{\chi\to \nu\gamma\gamma}^A
\approx
  9 \times 10^{-22}{\rm sec^{-1}}
\left(m_\chi \over 200\,\rm keV\right)^9
\left(\rm TeV \over \Lambda\right)^4,
\label{eq:chi2v2gAapp}
\end{eqnarray}
for tiny DM mass. Note that the dependence on
$m_\chi$ is higher by 2 powers than that of the
scalar and pseudo-scalar cases in \geqn{eq:chi2v2gapp}. 

Second, the invisible decay $\chi \to 3\nu$ shares the
similar features as that of the vector case including
the divergence. We use the same procedure to keep
only the leading non-divergent term,
\begin{eqnarray}
\Gamma_{\chi \to 3\nu}^{A}
\approx
{m_\chi^5 \over 1536\pi^3}
\left({3 g_2^2 \over 64\pi^2} \right)^2
{1 \over \Lambda^4}
\approx
3.66 \times 10^{-17}{\rm sec}^{-1} 
\left( m_\chi \over\rm 200~{\rm keV} \right)^5
\left( {\rm TeV} \over {\rm \Lambda} \right)^4.
\label{eq:chi23vA}
\end{eqnarray}
for the ${\cal O}_{e\nu\chi}^A$ operator with a muon/tau
neutrino. There is also an additional factor 2 for the
electron neutrino case.

\subsubsection{The Tensor Operator ${\cal O}_{e\nu\chi}^T$}

The dominant decay mode for the tensor operator is the
2-body decay $\chi \to \nu \gamma$ in \gfig{fig:chiDC}(a). 
However, it suffers from UV divergence, 
\begin{eqnarray}
  {\cal M}_{\chi\to\nu\gamma}
=
  {1 \over \Lambda^2}
  (\bar u_\nu \sigma^{\mu\nu}P_R u_\chi)
\left[
- i {e m_e q_\mu \epsilon^*_{\nu}\over 2\pi^2}
  \left({1\over \epsilon}+ \ln{\Lambda^2 \over m_e^2} \right) 
\right], 
\end{eqnarray}
and needs to be regularized by the DR scheme. The vectors $q_\mu$
and $\epsilon_\nu^*$ are the outgoing photon momentum
and polarization vector, respectively. After dropping
the divergent factor $1/\epsilon$, the decay width from
the the finite part becomes
\begin{eqnarray}
\Gamma_{\chi\to\nu\gamma}^T
=
{\alpha m_e^2 m_\chi^3  \over 16\pi^4 \Lambda^4}\ln^2{\Lambda^2 \over m_e^2}
\approx 1.5\times 10^{-8}{\rm sec^{-1}} 
\left(m_\chi \over 200\,\rm keV\right)^3
\left(\rm TeV \over \Lambda\right)^4
  {\ln^2 (\Lambda^2 / m_e^2) \over 1000},
\label{eq:chi2vga}
\end{eqnarray}
where the cut-off scale $\Lambda$ enters through the log term.
Since the logarithm is not sensitive to the change of $\Lambda$,
we approximate it by a typical value
$\ln^2 (\Lambda^2 / m_e^2) \sim {\cal O}(10^3)$. 
Comparing \geqn{eq:chi2vga} with \geqn{eq:chi2v2gapp},
\geqn{eq:chi2v3gapp} and \geqn{eq:chi2v2gAapp}, we see
that the $\chi \to \nu\gamma$ decay width for
${\cal O}_{e\nu\chi}^T$ is much larger than the dominant
visible decay width for all other operators. This implies
that a much stronger constraint will be put on the tensor operator. 
Again, the 3-body channel $\chi\to \nu\gamma\gamma$ vanishes
due to the QED charge conjugation symmetry . In addition,
the 4-body decay $\chi \to \nu\gamma\gamma\gamma$ is
suppressed by additional couplings as well as phase space
factor. 

For the invisible decay $\chi \to 3\nu$, as dictated by
Lorentz invariance and the left-handedness of SM neutrinos,
the electron tensor structure $(\bar e\sigma^{\mu\nu} e)$
induces an effective operator
$(m_e/m_W^2) \partial^\mu (\bar\nu_L \gamma^\nu \nu_L)$
after loop integration. The electron mass $m_e$ comes from
the chirality flip introduced by the tensor operator.
With the mass dimensions of $m_e$ and $\partial^\mu$
compensated by $1/m_W^2$, this contribution
is severely suppressed by a factor $m_e^2/ m_W^2\sim 10^{-11}$
than $\chi \to \nu \gamma$ and can be safely neglected.

\section{The cosmological and astrophysical constraints on DM Decays}
\label{sec:constraints}

As elaborated above, the DM absorption operators contain only
one DM field. There is no intrinsic mechanism to forbid DM
from decaying. This can provide some visible effect on the
cosmological evolution history and the astrophysical observations
via X-ray and gamma ray. This section evaluates first the
constraints from cosmology in \gsec{sec:cosmo} and astrophysical
observations in \gsec{sec:astro}.

\subsection{The Cosmological Evolution Constraints on the DM Invisible Decay $\chi\to 3\nu$}
\label{sec:cosmo}

As illustrated in \gsec{sec:DMDC}, both the vector and axial-vector
operators can have invisible decay $\chi \rightarrow 3 \nu$. More
importantly, the invisible decay mode dominates over the visible
ones by at least 4 orders. This can be seen by comparing
\geqn{eq:chi2v3gapp} with \geqn{eq:chi23vV}, and
\geqn{eq:chi2v2gAapp} with \geqn{eq:chi23vA}.
If a significant amount of
the DM decays invisibly to inject its energy into relativistic
degrees of freedom, the expansion history of the Universe can
receive sizable modifications. Previous studies have already
put quite strong constraints on the decaying DM scenario 
\cite{Gong:2008gi,DeLopeAmigo:2009dc,Audren:2014bca,Poulin:2016nat}.
The currently most stringent constraint is
$\Gamma_{\rm inv}^{-1} < 468\,\rm Gyr$ \cite{Abellan:2021bpx}. 

The constraints on $\sigma_{\chi e}^{V,A}v_\chi$ are shown as
dot-dashed lines in \gfig{fig:constraints}. The blue one is
for the vector case while the magenta one for the axial-vector
one. In the mass range
$40\,\rm keV \lesssim m_\chi \lesssim  500\,\rm keV$,
the constraints from $\chi\to 3\nu$ for the vector case is
stronger than the DM overproduction and gamma-ray constraints. 
Together with the approximation
$\sigma_{\chi e} v_\chi \approx m^2_\chi / 4 \pi \Lambda^4$,
the scaling behaviors
$\Gamma^{V,A}_{3 \nu} \propto m^5_\chi / \Lambda^4$
in \geqn{eq:chi23vV} and \geqn{eq:chi23vA} renders
the constraint to scale as $\propto 1 / m^3_\chi$. 
This estimation is consistent with the resulting curves
shown in \gfig{fig:constraints}. For other operators,
the $\chi \rightarrow 3 \nu$ channel is much smaller.

\subsection{The Astrophysical X-Ray and Gamma Ray Constraints on the Visible Decays $\chi\to \nu+\gamma(\rm s)$}
\label{sec:astro}

Although the visible decays are typically much smaller than the
invisible one as explored in \gsec{sec:DMDC}, it is much easier
to observe photon than neutrino. This is especially true in the
low energy range for sub-MeV DM. With DM distributing everywhere
in the Universe and being especially concentrated in our Milky Way galaxy,
the observation of diffuse X-ray and gamma ray can put stringent
constraints on the decay width and therefore the cut-off scale
$\Lambda$. We first describe how the DM visible decays contribute
to the X(gamma)-ray observations in \gsec{sec:xGammaRay}
and then compare with the astrophysics observation data sets in
\gsec{sec:astroData}.

\subsubsection{X-Ray and Gamma Ray Fluxes from the Visible Decays}
\label{sec:xGammaRay}

Both galactic and extra-galactic sources of DM visible decay
$\chi \rightarrow \nu \gamma (s)$ can
contribute to the X(gamma)-ray observations around our Earth.
Typically the extra-galactic contributions are much smaller than
the galactic counterpart. But for those diffuse cosmic fluxes,
the major contribution comes from extra-galactic sources. So we
will discuss both contributions below.

For DM decay, the galactic contribution is proportional to its local
density $\rho_\chi / m_\chi$ and the decay spectrum $d \Gamma_\chi / d E_\gamma$
calculated in the rest frame of DM.
So the differential photon flux per unit energy per solid angle is,
\begin{eqnarray}
  {d^2\Phi_\gamma \over dE_\gamma d\Omega} 
=
  \frac 1 {4 \pi} 
  \frac {d \Gamma_\chi} {d E_\gamma} 
  \int^{s_{\rm max}}_{\rm l.o.s} \frac {\rho_\chi(r)}{m_\chi} d s,
\end{eqnarray}
where $d \Gamma_\chi / d E_\gamma$ is the corresponding differential
decay width. The integration over the line of sight (l.o.s.) takes
all the contribution along a specific direction. Note that the DM
density $\rho_\chi(r)$ is a direct function of the distance $r$ from the
galactic center. We adopt the NFW profile, 
$\rho(r) = \rho_0 / [(r / r_s) (1 + r / r_s)^2]$ \cite{Navarro:1995iw,Navarro:1996gj},
where $r_s = 17$\,kpc \cite{Laha:2020ivk} and $\rho_0=0.43\,\rm GeV/cm^3$
to give the local DM density $\rho_\chi \approx 0.4\,\rm GeV/cm^3$.
The distance $r(s)$ is a function of the l.o.s. distance $s$ in the
galactic coordinate,
\begin{eqnarray}
  r(s)
=
  \sqrt{r_\odot^2 + s^2 -2 r_\odot s \cos\psi}.
\end{eqnarray}
In addition, $r_\odot = 8.3\,\rm kpc$ is the distance of Earth to
the galactic center and $\cos \psi \equiv \cos b\cos l$. 
The integration range for the l.o.s distance $s$ is from 0 to a
maximal value determined by the virial radius
$r_{\rm vir} = 300\,\rm kpc$ of the DM halo \cite{Lin:2019uvt}, 
\begin{eqnarray}
s_{\rm max} = r_\odot \cos\psi 
+ \sqrt{r_{\rm vir}^2 - r_\odot^2 \sin^2\psi }. 
\end{eqnarray}

The extra-galactic contribution comes from the smooth DM distribution
in the whole universe. Its contribution is isotropic
and integrated over a large range of the redshift \cite{Essig:2013goa},
\begin{eqnarray}
  \frac {d^2\Phi_{r}^{\rm EG}} {d E_\gamma d \Omega} 
=
  \frac {\Omega_{\rm DM} \rho_c}{4 \pi m_\chi 
  H_0\sqrt{\Omega_m}}
  \int_0^{\infty}
  \frac {d \Gamma_\chi}{d E_\gamma(z)}
  \frac {d z} {\sqrt{\kappa+ (1+z)^3}}.
\end{eqnarray}
The Hubble constant $H_0=67.4\,\rm km\,sec^{-1}Mpc^{-1}$
and the cosmological critical density
$\rho_c = 5.8\times 10^{-6}\,\rm GeV\,cm^{-3}$ are present
values. Of the total matter fraction $\Omega_m = 0.315$, DM
takes the largest share $\Omega_{\rm DM} = 0.265$. In addition,
the dark energy (DE) also has a large effect on the cosmological
evolution, especially in the late stage. We use
$\kappa \equiv {\Omega_\Lambda /\Omega_m}=2.17$ to parametrize
the contribution of DE. The decay width and spectrum calculated
in \gsec{sec:DMDC} cannot be used directly. Due to cosmological
redshift, the photon energy $E_\gamma(z) = (1 + z) E_\gamma$ emitted
at redshift $z$ is $1 + z$ times of the apparent $E_\gamma$.

The total photon flux ${d^2\Phi_{r} / dE_\gamma d\Omega}$ due
to DM decay is then a sum of the above two components. For a
telescope with effective area $A_{\rm eff}$ and field of view (FOV)
$\Delta\Omega$ as well as exposure time $T_{\rm obs}$,
the predicted photon event rate in energy bin $[E^-_i, E^+_i]$ is
\begin{eqnarray}
  N_i^{\rm th}
\equiv
  A_{\rm eff} T_{\rm obs} \int_{E^-_i}^{E^+_i} d E_\gamma 
  \int_{\Delta\Omega} d\Omega 
  {d^2\Phi_\gamma \over d E_\gamma d\Omega}.
\end{eqnarray}
Below we use real data to constrain the DM decay width and
subsequently the DM coupling strength.

\subsubsection{Constraints from Astrophysical X-ray and Gamma Ray Data in the keV-MeV Range}
\label{sec:astroData}

\begin{figure}[t]
\centering
\includegraphics[width=0.49\textwidth]{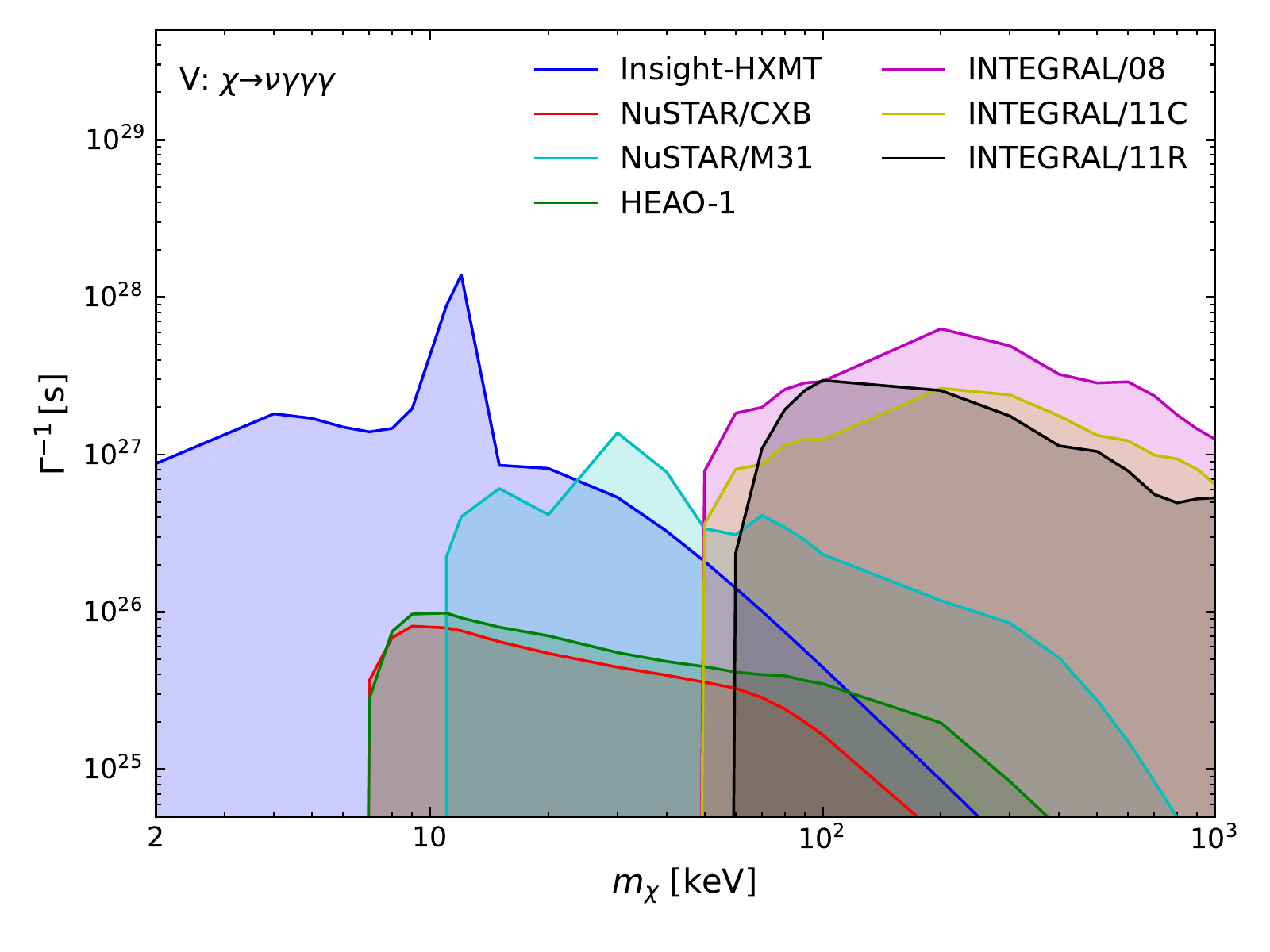}
\hfill
\includegraphics[width=0.49\textwidth]{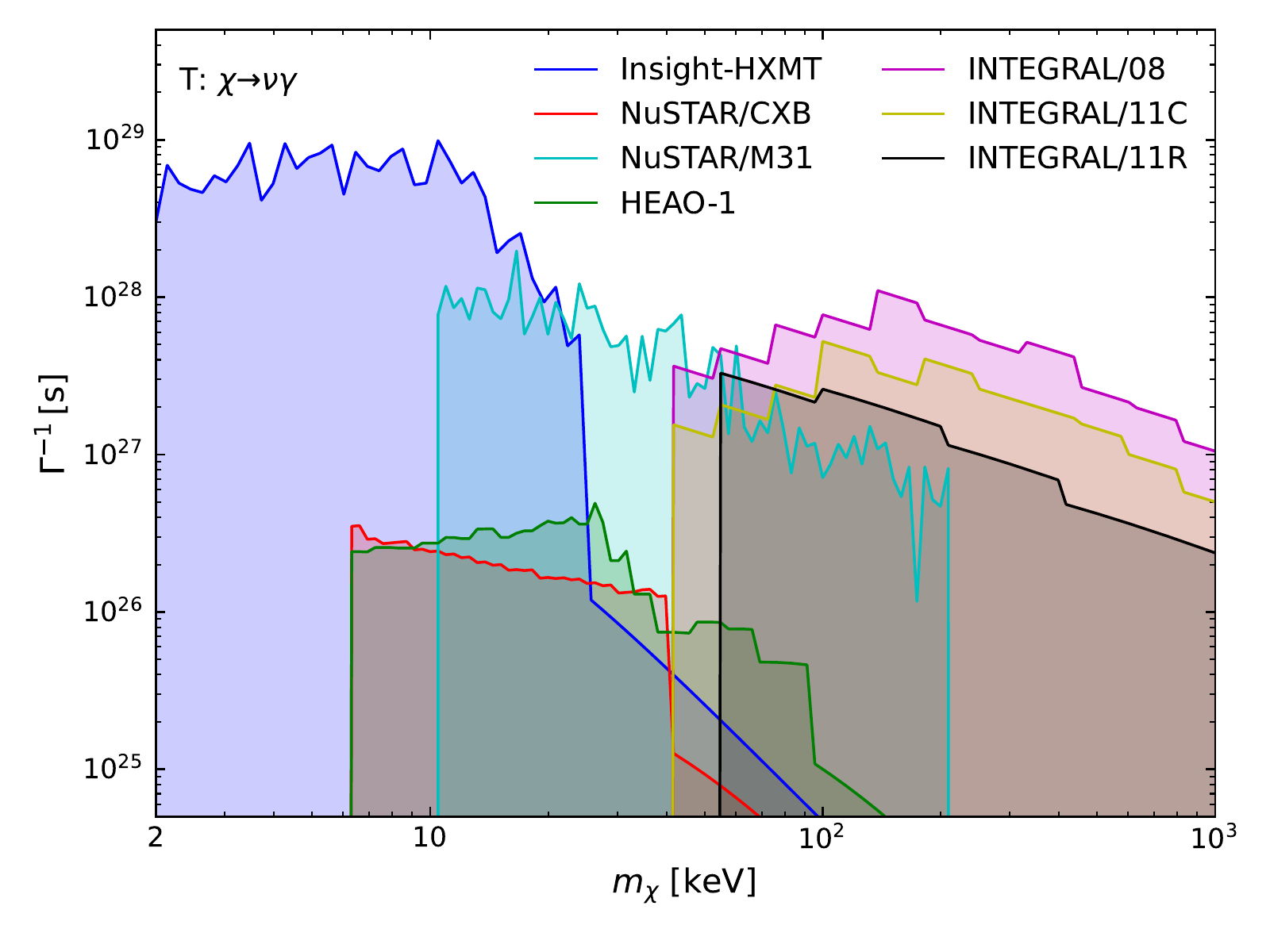}
\caption{The astrophysical X-ray and gamma-ray constraints on
the visible decay width of $\chi \rightarrow \nu + \gamma (s)$
as a function of the DM mass $m_\chi$. For illustration, the
vector operator with continuum spectrum from
$\chi \rightarrow \nu \gamma \gamma \gamma$ and the tensor
one with discrete $\delta$-function from
$\chi \rightarrow \nu \gamma$ are shown in the left and
right panels, respectively.}
\label{fig:astroDecayWidth}
\end{figure}

As argued at the beginning of \gsec{sec:FADM}, we are interested
in the DM mass range between keV and MeV. The relevant observations
in our analysis include Insight-HXMT \cite{Liao:2020hds},
NuSTAR \cite{Krivonos:2020qvl,Ng:2019gch}, HEAO-1 \cite{Gruber:1999yr},
and INTEGRAL \cite{Bouchet:2008rp,Bouchet:2011fn}. \gfig{fig:data}
summarizes the observed X(gamma)-ray data. Most data sets
are used to constrain the fermionic DM absorption operators for
the first time with the only exception of HEAO-1 and INTEGRAL-08
\cite{Essig:2013goa}. The constraints on the DM decay width
$\Gamma_\chi$ are shown in \gfig{fig:astroDecayWidth} while
the constraints on the interaction strength in terms of the
direct detection cross section $\sigma_{\chi e} v_\chi$ have
already been included in \gfig{fig:constraints} altogether.

To constrain the DM decay width $\Gamma_\chi$, we require
the predicted photon events in each energy bin does not
exceed the experimental counts at 95\% C.L. In a single energy bin
$[E^-_i, E^+_i]$, the constraint is obtained with
\begin{eqnarray}
  N_i^{\rm th} \leq N_i^{\rm obs} \equiv
  A_{\rm eff} T_{\rm obs} \Delta\Omega
\left( {d^2\Phi_\gamma \over dE_\gamma d\Omega} \right)_{\rm exp@95\%}^i
  \Delta E_i.
\end{eqnarray}
In principle, one may directly compare the predicted flux
$d^2 \Phi_r / d E_\gamma d \Omega$ with the data in
\gfig{fig:data} without converting to event number in each
bin. Nevertheless, the spectrum of the two-body channel
$\chi \rightarrow \nu \gamma$ for the tensor operator
$\mathcal O^T_{e \nu \chi}$ is a $\delta$ function which
is difficult to directly compare with \gfig{fig:data}.
With multiple data points,
we can obtain a corresponding limit for the decay width
$\Gamma_\chi^i$ from the $i$-th energy bin, and we take the
strongest bound among all bins as the final limit for the
corresponding mass point. 

Some data releases, especially Insight-HXMT \cite{Liao:2020hds}
and NuSTAR/M31 \cite{Ng:2019gch},
even provide background models in addition to data points.
This opens the possibility to use $\chi^2$ fit to
obtain enhanced sensitivity than simply comparing with the
central value plus the 95\% C.L. uncertainty for individual bins.
Putting things together, the corresponding $\chi^2$ function for
fitting the NuSTAR/M31 data is,
\begin{eqnarray}
  \chi^2(x_i, \Lambda)
\equiv
  \sum_i
\left[
  \frac {\Lambda^{-4}_{\rm TeV} N_i^{\rm DM}({\Lambda = 1\,\rm TeV})
       + \sum_a c_a N^a_i
       - N_i^{\rm exp}}
        {\delta N_i}
\right]^2\!\!\!,
\end{eqnarray}
where $\Lambda_{\rm TeV} \equiv \Lambda / \rm TeV$.
The observation data provides the central values $N^{\rm exp}_i$
and the corresponding uncertainty $\delta N_i$ for the $i$-th bin.
Each observation can have multiple backgrounds $N^a_i$ with
$a$ denoting its type and $c_a$ the corresponding normalization factor.
The $\chi^2$ fit with data takes $c_a$ as fitting parameters
while the one for the DM contribution is the cut-off
scale $\Lambda_{\rm TeV}$ in unit of TeV. More details of
the analytic $\chi^2$ fit can be found in \gapp{app:chi2fit}.

Below is a detailed description of each data set and their constraints
on the DM decay width and coupling strength.

\begin{itemize}
\item {\bf Insight-HXMT/CXB}: We use the $(1 \sim 12)$\,keV
cosmic X-ray background (CXB) data observed by the Low Energy
X-ray Telescope on Insight-HXMT (Hard X-ray Modulation Telescope)
\cite{Liao:2020hds}. The observation points to the sky in the 
direction $(l, b)=(219.3^\circ, -50.0^\circ)$ with a small FOV
($1.6^\circ\times 6^\circ$). The relevant effective detector area
is taken from the Fig.\,1 on the HXMT website \cite{hxmt-web}.
For comparison, the background model for CXB is taken from the
yellow line of the Fig.\,11 therein. With both data points and
background model provided, we use analytic $\chi^2$ fit to
obtain constraint. The result is shown in \gfig{fig:astroDecayWidth}
with blue color. In the $\mathcal O(1)$\,keV range,
Insight-HXMT/CXB gives a strong constraint.     

\begin{figure}
\centering
\includegraphics[width=14cm]{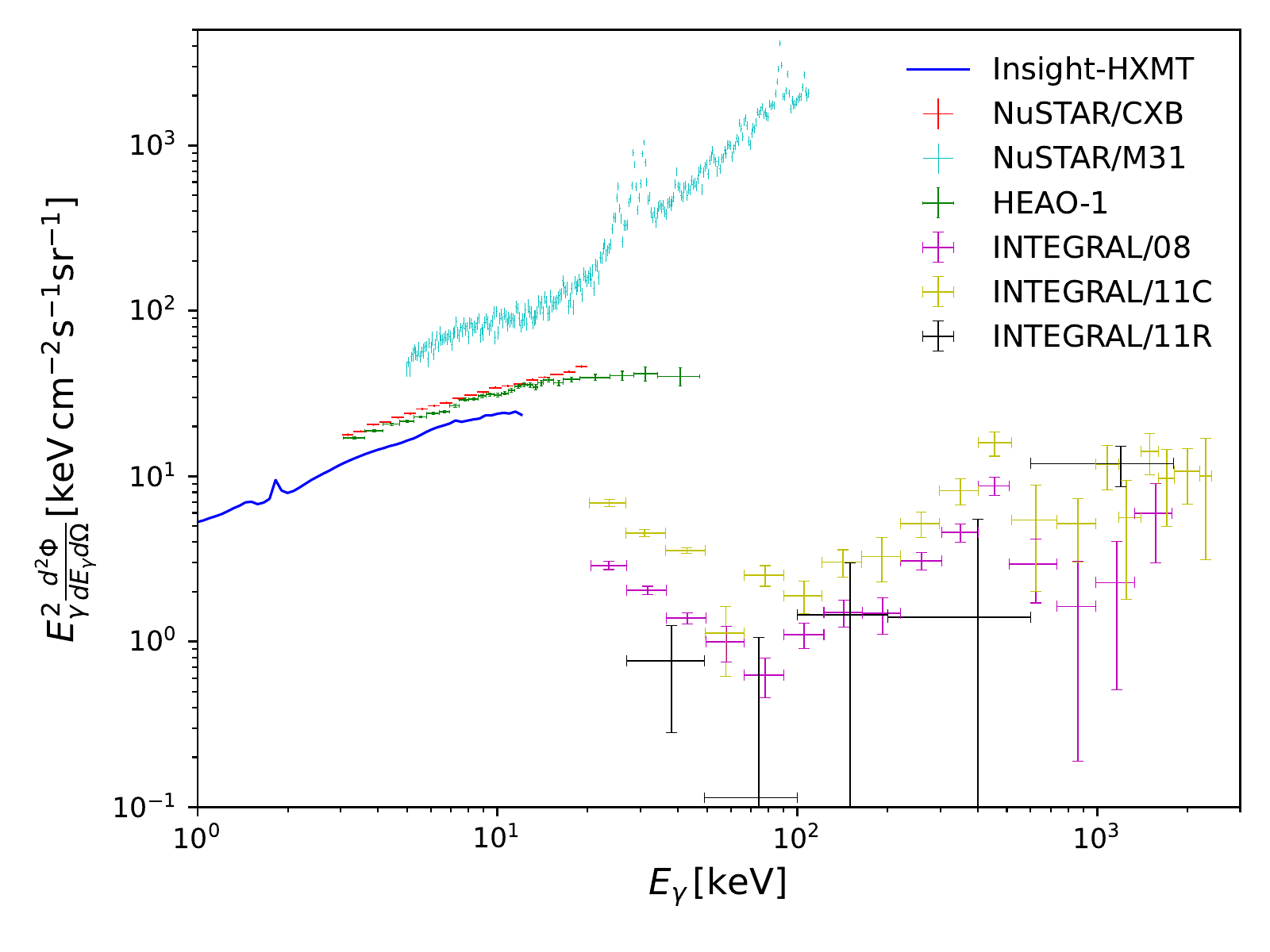}
\caption{The observed X(gamma)-ray fluxes by Insight-HXMT/CXB
(blue) \cite{Liao:2020hds}, NuSTAR/CMB (red)
\cite{Krivonos:2020qvl}, NuSTAR/M31 (cyan)
\cite{Ng:2019gch}, HEOA-1 (green) \cite{Gruber:1999yr},
INTEGRAL (magenta, yellow, and black)
\cite{Bouchet:2008rp,Bouchet:2011fn}. 
The three INTEGRAL data sets are released in 2008 (magenta)
\cite{Bouchet:2008rp} and 2011 (yellow for 11C and black for 11R)
\cite{Bouchet:2011fn} for different observational sky regions
as described in the main text.}
\label{fig:data}
\end{figure}

\item {\bf NuSTAR/CXB}: The Fig.\,10 of \cite{Krivonos:2020qvl}
gives the average CXB spectrum within the $3 \sim 20$\,keV energy
range. This spectrum is obtained by
stacking the focal plane module (FPM) A and B (FPMA and FPMB)
observations with all six data sets (COSMOS EP1,2,3, EGS, ECDFS, UDS).
 
\item {\bf NuSTAR/M31}: In addition to the diffuse CXB, galaxy
observation can also provide a strong constraint due to the
concentrated DM density profile. A typical case is
the NuSTAR observation of M31. We use the $(5 \sim 100)$\,keV
data in the Fig.\,2 of \cite{Ng:2019gch} from the observation
ID 50026002003. The NuSTAR instrumental and solar contributions,
the 0-bounce CXB component, and the 2-bounce component from the
diffuse M31 emission are taken into consideration as backgrounds.
The background models are taken from the fit curves in the
Fig.\,2 of \cite{Ng:2019gch}. Each
component has its own normalization factor as fitting parameter.
Since the 2-bounce CXB component is very small, we neglect
it in our $\chi^2$ fit to avoid numerical instability.

For the DM decay photons, the
0-bounce and 2-bounce DM decay photons have different effective
areas. To properly take the 2-bounce contribution into account,
we use the enhancement factor defined in \cite{Ng:2019gch},
\begin{eqnarray}
  \xi(E_\gamma)
\equiv
  1
+ \frac {A_{\rm 2b}(E_\gamma) \Delta\Omega_{\rm 2b} {\cal J}_{\rm 2b}}
        {A_{\rm 0b} \Delta\Omega_{\rm 0b} {\cal J}_{\rm 0b}}.
\end{eqnarray}
Then the predicted DM decay photon events in each energy
bin can be written as,
\begin{eqnarray}
  N_i^{\rm DM}
\equiv
  A_{\rm 0b} T_{\rm obs} \Delta\Omega_{\rm 0b}
  \frac {\mathcal J_{\rm 0b}} {4 \pi m_\chi} 
  \int d E_\gamma \xi(E_\gamma)  {d \Gamma_\gamma \over d E_\gamma } ,
\end{eqnarray}
where the observational effective area for the 0-bounce photons
is $A_{0b}= 11.85 \, (11.80)\,\rm cm^2$, the exposure time is
$T_{\rm obs}= 82.4 \, (82.2)\,\rm ks$ and the FOV
$\Delta\Omega_{0b}= 4.45 \, (4.55) \,\rm deg^2$ for the FPMA (FPMB)
observation, respectively. The DM decay factor
${\cal J}_{\rm 0b} = 6.72\,(7.13)\, \rm GeV\,cm^{-3}\,kpc\,sr^{-1}$
for FPMA (FPMB) includes both the Milky Way and M31 contributions.
One can neglect the extragalactic contribution which is
much smaller.
 
\item {\bf HEOA-1}: The HEAO-1 extragalactic diffuse
X-ray data in the Fig.\,2 of \cite{Gruber:1999yr}
corresponds to the sky region $l\in (58^\circ, 106^\circ)
\cup (238^\circ,289^\circ), |b|\in(20^\circ, 90^\circ)$.
Following \cite{Essig:2013goa}, we only use the $(3 \sim 50)$\,keV
data set observed by the A2 High-Energy Detector (HED).
 
\item {\bf INTEGRAL}: We use both the galactic center
gamma-ray spectrum ($|b|< 15^{\circ}$ and $|l| < 30^{\circ} $)
as well as the galactic ridge emission spectrum from
the SPI measurements on board INTEGRAL. 1) There are two data
sets for the galactic center gamma-ray spectrum with
photon energy $(20 \sim 2000)$\,keV. One is from the
Fig.\,9 of \cite{Bouchet:2008rp} released in 2008 and
shown as INTEGRAL/08 (magenta) in \gfig{fig:data} while
the other comes
from the Fig.\,6 of \cite{Bouchet:2011fn} released in
2011 and shown as INTEGRAL/11C (yellow). 2) For the galactic ridge
emission spectrum, the Fig.\,4 of \cite{Bouchet:2011fn} gives
the diffuse emission as a function of galactic longitude
with the latitude being integrated over and the Fig.\,5 therein
gives the one with the longitude being integrated over.
The INTEGRAL measurements from this analysis are divided
into five energy bins with divisions at $E=(27, 49, 100, 200, 600,
1800)$\,keV, respectively. We select those bins that
give the strongest limit and show their corresponding
fluxes in \gfig{fig:data} as INTEGRAL/11R (black).

\end{itemize}

\gfig{fig:astroDecayWidth} compiles all the astrophysical
X-ray and gamma-ray constraints on the DM visible decay.
While the photon energy is typically smaller than half
of the DM mass, $E_\gamma < m_\chi/2$, the energy range
$(1 \sim 3000)$\,keV in \gfig{fig:data} covers
the DM mass window $(2 \sim 1000)$\,keV in
\gfig{fig:astroDecayWidth}.
Although there are three different decay channels,
$\chi \rightarrow (\nu \gamma, \nu \gamma \gamma,
\nu \gamma \gamma \gamma)$, the last two share similar
features of continum spectrum while the first has a
discrete $\delta$ function. For illustration purpose,
we only show the vector and tensor cases in the left
and right panels of \gfig{fig:astroDecayWidth}, respectively.
Due to this difference, the curves for the vector case are
quite smooth while the tensor ones have many breaks and
spikes. In addition, the tensor case typically has much
clearer boundaries such as the NuSTAR/M31 curve.
The results for $\chi \rightarrow \nu \gamma \gamma$ are
quite similar to those of
$\chi \rightarrow \nu \gamma \gamma \gamma$.

It is interesting to observe that, although the NuSTAR/M31
flux in \gfig{fig:data} is not as small as other observations,
its constraint on the DM visible decay width is not bad and
even better than some others such as INTEGRAL. This is because the INTEGRAL
constraints comes from comparing the theoretical prediction
with all observed event counts plus errors at 95\% C.L. while
the NuSTAR/M31 constraint is comes from a more realistic
$\chi^2$ fit. If possible, $\chi^2$ fit is more desirable
although doing this for all astrophysical data is beyond the
scope of the current paper.

Another important feature is that, the constraining power
can go beyond the $E_\gamma < m_\chi / 2$ correspondence.
Taking the Insight-HXMT curve for demonstration, the
adopted spectrum spans the energy range $(1 \sim 12)$\,keV
while the constrained mass range can extend up to
$\mathcal O(100)$\,keV. This is because the extragalactic
contributions from the vast Universe receive redshift to
different extent. Although the emitted photon spectrum is
fixed by the DM mass $m_\chi$ and the decay vertex, the
observed photon energy could be much lower. A heavier
$m_\chi$ above the energy window can also receive constraint
from low energy X-ray observation.

The constraints in terms of the direct detection cros section
$\sigma_{\chi e} v_\chi$ have already been shown in
\gfig{fig:constraints} for comparison. Comparing with the
overproduction constraints, the decay constraints are typically
more stringent for heavier DM for both the invisible and
visible channels. This is because the decay width typically
grows with the DM mass. The constraint for the tensor operator
is particularly strong since it comes from the single photon
channel $\chi \rightarrow \nu \gamma$ with much larger phase
space. The next highly constrained operator is the pseudo-scalar
type. Neither the freeze-in production nor decay process of the
pseudo-scalar operator is suppressed for the others. So the
constraints on its cut-off scale should be roughly the same
as others. However, the direct detection cross section is
highly suppressed for the pseudo-scalar case. Both pseudo-scalar
and tensor operators are difficult to be directly probed.
The constraints on the other operators are not that severe.
Of them, the vector case is of particular interest for
tonne-scale direct detection experiments
(such as PandaX-4T \cite{PandaX:2018wtu,PandaX-4T:2021bab},
XENONnT \cite{XENON:2020kmp},
and LZ \cite{LZ:2015kxe,Mount:2017qzi}) which will
soon be able to probe small DM mass $m_\chi$
of $\mathcal O(10 \sim 100)$\,keV that has not been
excluded by cosmological or astrophysical constraints.

\section{Conclusions}
\label{sec:summary}

We systematically investigated the
fermionic DM absorption on the electron target
that allow unique probe of sub-MeV DM. Using the effective
fermionic absorption operators, we found that the electron
recoil spectrum in direct detection has roughly the same
shape that is mainly determined by the atomic $K$-factor
for different operators. This allows a model-independent,
or at least operator-independent, measurement to some extent.
It even allows in-situ measurement of the atomic $K$-factor
if the fermionic DM absorption is confirmed. The only
complication is that the pseudo-scalar case has quite
different signal size. The comparison with the
Xenon1T and PandaX-II electron recoil spectrum prefers a vector-type
DM absorption with $m_\chi = 59$\,keV and 105\,keV respectively.
With the corresponding best-fit value $\Lambda \approx 1$\,TeV,
the Xenon1T and PandaX-II can probe the new physics cut-off scale
up to TeV scale. We also systematically update the
overproduction, cosmological, and astrophysical constraints.
Especially, the X(gamma)-ray constraints from the Insight-HXMT,
NuSTAR, and INTEGRAL 2011 data sets are newly used to constrain
the sub-MeV fermionic absorption DM. Even though the tensor
and pseudo-scalar operators are strongly constrained, the
fermionic DM absorption with other operator types is still
testable at tonne-scale experiments.

\section*{Acknowledgements}

The authors would like to thank
Roman Krivonos,
Lei Lei, 
Jin-Yuan Liao, 
Jiang-Lai Liu, 
Dan Zhang, and
Shuang-Nan Zhang for useful discussions.
The authors thank Kenny C. Y. Ng for providing us the NuSTAR/M31 data,
the ${\cal J}$ factor and the enhancement factor of the 2 bounce FOV in
their paper \cite{Ng:2019gch}. The authors also thank Jeff A. Dror for
double-checking the results in \cite{Dror:2020czw}.
This work is supported in part by the Double First Class start-up fund
(WF220442604), the Shanghai Pujiang Program (20PJ1407800),
National Natural Science Foundation of China (Nos. 12090064, 11975149, 11735010),
Chinese Academy of
Sciences Center for Excellence in Particle Physics (CCEPP), and Key Laboratory for Particle Physics, Astrophysics and Cosmology, Ministry of Education, and Shanghai Key Laboratory for Particle Physics and Cosmology (Grant No. 15DZ2272100). XGH was also supported in part by the MOST (Grant No. MOST 106- 2112-M-002-003-MY3 ).

\appendix

\section{Analytic $\chi^2$ Fit with Collective Marginalization}
\label{app:chi2fit}

The fitting with experimental data points in this paper is achieved
with analytical $\chi^2$ fit \cite{Ge:2012wj,Ge:2016zro}. With
Gaussian distribution, the $\chi^2$ minimization is equivalent to
matrix manipulation. Most importantly, the marginalization for
a single parameter can also be done as matrix element manipulation
to reduce a $\chi^2$ function with $n$ parameters to the one with
$n-1$ parameters. This single-parameter marginalization needs
to be done recursively in order to marginalize over multiple
parameters. Here we provide a more elegant formalism to
marginalize over multiple parameters collectively.

Given a set of observables ${\cal O}_j$, the $\chi^2$ function
can be generally parametrized as,
\begin{eqnarray}
  \chi^2
=
  \sum_j
\left({{\cal O}_j^{\rm th} -{\cal O}_j^{\rm exp} \over \Delta {\cal O}_j } \right)^2, 
\end{eqnarray}
where ${\cal O}_j^{\rm th}$ and ${\cal O}_j^{\rm exp}$ are the
theoretical prediction and experimental observation for the
$j$-th bin, respectively.
A Gaussian $\chi^2$ function is equivalent to linear dependence
of ${\cal O}_j^{\rm th}$ on the model parameters $x_i, (i=1,2,...,n)$
as ${\cal O}_j^{\rm th} \approx \mathcal O^{\rm th,0}_j + \sum_i A_{ji} x_i$.
In matrix form, the $\chi^2$ function can be written as 
\begin{eqnarray}
  \chi^2(x_i)
=
  (\mathcal O^{\rm th,0} + A x-{\cal O}^{\rm exp})^{\rm T}
  \overline \Sigma^{-1}
  (\mathcal O^{\rm th,0} + A x-{\cal O}^{\rm exp}), 
\quad
  \overline \Sigma^{-1}
\equiv
  {\rm diag}(\Delta {\cal O}_j^{-2}).
\label{eq:chi2}
\end{eqnarray}
For $m$ observales and $n$ fitting parameters, $A$ is a $m \times n$
constant coefficient matrix and $x \equiv (x_1,\cdots, x_n)^T$
is a $n \times 1$ column vector. Then $A$ converts the $n \times 1$
parameter vector $x$ to a $m \times 1$ observable vector $A x$ that can
match with $\mathcal O^{\rm th,0}$ and $\mathcal O^{\rm exp}$. Finally,
the $m \times m$ error matrix $\overline \Sigma^{-1}$ in the observable
space contracts with two observable vectors, one column and
one row vectors, to produce a scalar $\chi^2$ function. By definition,
error matrix is symmetric.

The $\chi^2$ minimization condition $\partial \chi^2/\partial x_i\equiv 0$
gives a unique solution for the best fit value of the fitting parameters,
\begin{eqnarray}
  x_{\rm best}
\equiv
  (A^T \overline \Sigma^{-1} A)^{-1}
  A^T \overline \Sigma^{-1}
  (\mathcal O^{\rm th,0} - {\cal O}^{\rm exp}).
\end{eqnarray}
With larger deviation between the experimentally observed
$\mathcal O^{\rm exp}$ and the zeroth-order prediction
$\mathcal O^{\rm th,0}$, the fitting parameter should also
deviate more from the one used to predict $\mathcal O^{\rm th,0}$.
Correspondingly, the $\chi^2$ function splits into two parts
\begin{eqnarray}
  \chi^2(x_i)
=
  \chi^2_{\rm min}
+ (x-x_{\rm best})^T \Sigma^{-1} (x-x_{\rm best}), 
\label{eq:chi2x}
\end{eqnarray}
Now $\chi^2$ becomes a function of fitting parameters $x_i$,
instead of observables, with the corresponding $n \times n$ error matrix
$\Sigma^{-1} \equiv A^T \overline \Sigma^{-1} A$ in the parameter space.
The first term of \geqn{eq:chi2x} is the minimum value of the $\chi^2$, 
\begin{eqnarray}
  \chi^2_{\rm min}
\equiv
  (\mathcal O^{\rm th,0} - {\cal O}^{\rm exp})^T
  B^{\rm T} \overline \Sigma^{-1} B
  (\mathcal O^{\rm th,0} - {\cal O}^{\rm exp})^T,
\quad
  B
\equiv
  \mathbb{I}
- A (A^T \overline \Sigma^{-1} A)^{-1} A^T \overline \Sigma^{-1},
\end{eqnarray}
while the second is actually $\delta \chi^2(x_i)$ as deviation from
$\chi^2_{\rm min}$.

With multiple fitting parameters in \geqn{eq:chi2x}, it is difficult
to see the probability distribution of any specific one. It is
desirable to obtain the $\chi^2$ of a single parameter by marginalizing
over the others. This can be achieved by integrating out the unnecessary
ones from the distribution function $\mathbb P(x_1, \cdots x_n)$.
Taking one-parameter reduction for illustration,
\begin{eqnarray}
  \mathbb P(x_1, \cdots, \hat x_k, \cdots, x_n)
=
  \int \mathbb P(x_1, \cdots, x_n) d x_k,
\end{eqnarray}
where the $k$-th element is marginalized. For a Gaussian distribution,
this is equivalent to matrix element manipulation of the error matrix
$\Sigma^{-1}$ in the parameter space,
\begin{eqnarray}
  \widetilde \Sigma^{-1}_{ij}
=
  \Sigma^{-1}_{ij}
- \frac {\Sigma^{-1}_{ik} \Sigma^{-1}_{jk}}
        {\Sigma^{-1}_{kk}}.
\end{eqnarray}
While $\Sigma^{-1}$ being a $n \times n$ matrix, $\widetilde \Sigma^{-1}$
is $(n-1) \times (n-1)$ after marginalizing one single parameter $x_k$.
Keeping doing this repeatedly, one can finally arrive at a $\chi^2$
function with only one parameter.

Nevertheless, this procedure is a little bit troublesome with $n-1$
repetition when $n$ becomes large. Below we provide a more convenient
algorithm of collective marginalization which can reduce a $n$-parameter
$\chi^2$ function directly to a single-parameter one without repetition.
Suppose one needs to marginalize over $k$ parameters out of the original
$n$ ones. Instead of using a single $n \times 1$ vector $x$, the fitting
parameters can be separated into one $(n-k) \times 1$ vector $X$ that
shall remain and one $k \times 1$ vector $Y$ that needs to be marginalized
away. Correspondingly, the experimental observables is predicted as
\begin{eqnarray}
  \mathcal O^{\rm th,0} + A_X X + A_Y Y,
\end{eqnarray}
instead of the original $\mathcal O^{\rm th,0} + A x$. The $\chi^2$
function \geqn{eq:chi2} then becomes
\begin{eqnarray}
  \chi^2(X,Y)
=
  (\mathcal O^{\rm th,0} + A_X X + A_Y Y - {\cal O}^{\rm exp})^T
  \overline \Sigma^{-1}
  (\mathcal O^{\rm th,0} + A_X X + A_Y Y - {\cal O}^{\rm exp}).
\end{eqnarray}
For convenience, one may define
$\delta \mathcal O_X \equiv \mathcal O^{\rm th,0} + A_X X - \mathcal O^{\rm exp}$
and the $\chi^2(X,Y)$ function becomes,
\begin{eqnarray}
  \chi^2(X, Y)
=
  Y^T \Sigma^{-1}_Y Y
+ 2 Y^T A^T_Y \overline \Sigma^{-1} \delta \mathcal O_X
+ \delta \mathcal O^T_X \overline \Sigma^{-1} \delta \mathcal O_X,
\label{eq:chi2XY}
\end{eqnarray}
with $\Sigma^{-1}_Y \equiv A^T_Y \overline \Sigma^{-1} A_Y$.
Only the first two terms are relevant in the Gaussian integration
of the marginalization of $Y$,
\begin{eqnarray}
  \mathbb P(X)
=
  \int \mathbb P(X,Y) d Y
=
  N \int e^{- \frac 1 2 \chi^2(X, Y)} d Y,
\label{eq:GaussianInteg}
\end{eqnarray}
with a normalization factor $N$ that would not affect the probability
distribution. The result would be more transparent by reforming
\geqn{eq:chi2XY} as,
\begin{eqnarray}
  \chi^2(X, Y)
& = &
  (Y + \Sigma_Y A^T_Y \overline \Sigma^{-1} \delta \mathcal O_X)^T
  \Sigma^{-1}_Y
  (Y + \Sigma_Y A^T_Y \overline \Sigma^{-1} \delta \mathcal O_X)
\nonumber
\\
& + &
  \delta \mathcal O^T_X
\left[
  \overline \Sigma^{-1}
- \overline \Sigma^{-1} A_Y \Sigma_Y A^T_Y \overline \Sigma^{-1}
\right]
  \delta \mathcal O_X.
\end{eqnarray}
The first line is a Gaussian form of $Y$ while the second
line is independent of $Y$. Then the Gaussian integration
\geqn{eq:GaussianInteg} gives,
\begin{eqnarray}
  \mathbb P(X)
\propto
  \exp
\left[
- \frac 1 2
  (\mathcal O^{\rm th,0} + A_X X - {\cal O}^{\rm exp})^T
  \overline \Sigma^{-1}_X
  (\mathcal O^{\rm th,0} + A_X X - {\cal O}^{\rm exp})
\right],
\label{eq:PX}
\end{eqnarray}
up to a normalization factor. Since a Gaussian probability
distribution is defined as $\mathbb P(X) \propto e^{- \chi^2(X)/2}$,
one can read off the reduced $\chi^2(X)$ directly from the
above equation. The reduced experimental error matrix,
\begin{eqnarray}
  \overline \Sigma^{-1}_X
\equiv
  \overline \Sigma^{-1}
- \overline \Sigma^{-1} A_Y \Sigma_Y A^T_Y \overline \Sigma^{-1},
\end{eqnarray}
with only parameters $X$ replaces the original $\overline \Sigma^{-1}$.
It is interesting to observe that the reduced $\chi^2(X)$
in \geqn{eq:PX} resembles the original form \geqn{eq:chi2}.
The effect of $Y$ parameters is wholely encoded in
$\overline \Sigma^{-1}_X$. It not only affects the $X$
error matrix,
$\Sigma^{-1}_X \equiv A^T_X \overline \Sigma^{-1}_X A_X$,
but also its best fit values. The marginalization down to
a single parameter corresponds to $k = n-1$.

\end{document}